\documentclass[fleqn,usenatbib]{mnras}

\usepackage{newtxtext,newtxmath}

\usepackage[T1]{fontenc}
\usepackage{ae,aecompl}

\usepackage{graphicx}	
\usepackage{amsmath}	
\usepackage{amssymb}	
\usepackage{hyperref}

\newcommand*\diff{\mathop{}\!\mathrm{d}} 

\def\Mv{M_{\rm v}}
\def\Ms{M_{\rm s}}
\def\Mg{M_{\rm g}}

\def\Rd{R_{\rm d}}
\def\Hd{H_{\rm d}}
\def\Rv{R_{\rm v}}
\newcommand{\msun}{M_\odot}
\newcommand{\yr}{\,{\rm yr}}
\newcommand{\pc}{\,{\rm pc}}
\newcommand{\cmc}{\,{\rm cm}^{-3}}

\title[Evaluating Galaxy Dynamical Masses]{Evaluating Galaxy Dynamical Masses From Kinematics and Jeans Equilibrium in Simulations}

\author[M. Kretschmer et al.]{Michael Kretschmer,$^{1,2}$\thanks{E-mail: michael.kretschmer@physik.uzh.ch}
Avishai Dekel,$^{2}$
Jonathan Freundlich,$^{2,3}$
Sharon Lapiner,$^{2}$
\newauthor Daniel Ceverino,$^{4,5}$
Joel Primack$^{6}$
\\
$^{1}$Institute for Computational Science, University of Zurich, Winterthurerstrasse 190, 8057 Zurich, Switzerland\\
$^{2}$Center for Astrophysics and Planetary Science, Racah Institute of Physics, The Hebrew University, Jerusalem 91904, Israel\\
$^{3}$School of Physics and Astronomy, Tel Aviv University, Tel Aviv 69978, Israel\\
$^{4}$Departamento de Fisica Teorica, Modulo 8, Facultad de Ciencias, Universidad Autonoma de Madrid, 28049 Madrid, Spain\\
$^{5}$CIAFF, Facultad de Ciencias, Universidad Autonoma de Madrid, 28049 Madrid, Spain\\
$^{6}$Physics Department, University of California, Santa Cruz, 1156 High Street, Santa Cruz, CA 95064, USA
}

\date{Accepted XXX. Received YYY; in original form ZZZ}

\pubyear{2020}

\begin{document}
\label{firstpage}
\pagerange{\pageref{firstpage}--\pageref{lastpage}}
\maketitle

\begin{abstract}
We provide prescriptions to evaluate the dynamical mass ($M_{\rm dyn}$) of galaxies from kinematic measurements of stars or gas using analytic considerations and the VELA suite of cosmological zoom-in simulations at $z=1-5$.
We find that Jeans or hydrostatic equilibrium is approximately valid for galaxies of stellar masses above $M_\star \!\sim\! 10^{9.5}M_\odot$ out to $5$ effective radii ($R_e$).
When both measurements of the rotation velocity $v_\phi$ and of the radial velocity dispersion $\sigma_r$ are available, the dynamical mass $M_{\rm dyn} \!\simeq\! G^{-1} V_c^2 r$ can be evaluated from the Jeans equation $V_c^2= v_\phi^2 + \alpha \sigma_r^2$ assuming cylindrical symmetry and a constant, isotropic $\sigma_r$.
For spheroids, $\alpha$ is inversely proportional to the Sérsic index $n$ and $\alpha \simeq 2.5$ within $R_\mathrm{e,stars}$ for the simulated galaxies.
The prediction for a self-gravitating exponential disc, $\alpha = 3.36(r/R_e)$, is invalid in the simulations, where the dominant spheroid causes a weaker gradient from $\alpha \!\simeq\! 1$ at $R_\mathrm{e,gas}$ to 4 at $5R_\mathrm{e,gas}$.
The correction in $\alpha$ for the stars due to the gradient in $\sigma_r(r)$ is roughly balanced by the effect of the aspherical potential, while the effect of anisotropy is negligible.
When only the effective projected velocity dispersion $\sigma_l$ is available, the dynamical mass can be evaluated as $M_{\rm dyn} = K G^{-1} R_e \sigma_l^2$, where the virial factor $K$ is derived from $\alpha$ given the inclination and $v_\phi/\sigma_r$.
We find that the standard value $K=5$ is approximately valid only when averaged over inclinations and for compact and thick discs, as it ranges from 4.5 to above 10 between edge-on and face-on projections.
\end{abstract}




\section{Introduction}
Estimating the total mass of a galaxy is an important challenge in astrophysics, in particular to assess the dark matter (DM) fraction within given radii and its potential evolution with cosmic time. Recent kinematic observations by \citet{2016ApJ...831..149W} and \citet{2017Natur.543..397G,2020ApJ...902...98G} indicate low central DM fractions in massive star-forming disc galaxies at $z=0.6-2.6$ within their effective radius. If confirmed, this result would have important implications for our understanding of galaxy formation and evolution, since it requires to rapidly drive DM out of the initial central cusps while baryons move inwards. Mechanisms such as dynamical friction \citep[e.g.][]{2001ApJ...560..636E,2004ApJ...607L..75E} and feedback from stars and AGN \citep[e.g.][]{2012MNRAS.421.3464P,2016MNRAS.461.1745E,2020MNRAS.491.4523F} or the combination of both may account for the observed low DM fractions in the early universe (Dekel et al., in prep.). More generally, estimating the total mass of a galaxy enables us to better understand the interplay between baryons and DM and hence to test the predictions of the LCDM model of structure formation.

Using kinematic data, it is possible to infer the dynamical mass from simple mass-estimation models based on Jeans equilibrium. 
However, the derived total mass from observed kinematics is uncertain and sometimes even turns out to be smaller than the stellar mass, indicating that the method used is flawed. It is therefore crucial to verify the validity of the Jeans equation and to quantify the associated expression for mass estimation during the different phases of galaxy evolution and for the different components -- spheroid and disc, stars and gas.

Observationally, galaxy kinematics can be derived from spatially-resolved emission line maps.
To measure gas rotation and dispersion profiles, position-velocity cuts along the galaxy's major axis are fitted assuming Gaussian line profiles, such that both the rotation curve and the velocity dispersion are measured simultaneously \citep{2008MNRAS.385..553D,2007ApJ...660L..35K,2012ApJ...758..106K,2019MNRAS.487.2924B,2016ApJ...831..149W,2014MNRAS.444.3894H}.

When both, the rotation velocity $v_\phi$ and velocity dispersion $\sigma_r$ are available, the dynamical mass can be assessed from the Jeans equation, which implies for cylindrically symmetric systems with a constant radial velocity dispersion \citep[][Section 4.8]{1915MNRAS..76...70J,2008gady.book.....B}:
\begin{equation}
V_c^2(r) = v_\phi^2(r) + \alpha(r) \sigma_r^2
\label{eq1}
\end{equation}
where $V_c$ is the circular velocity associated with the gravitational potential and $\alpha$ is a dimensionless parameter.
The right hand side is valid for each component of the galaxy separately, disc or bulge, gas or stars, potentially with different $\alpha(r)$.
One needs estimates of $\alpha(r)$ for the different galactic components, either derived analytically in simple cases or computed from simulations.

Observationally it is often difficult to obtain independent reliable values for rotation and dispersion such that only a line-of-sight (los) velocity dispersion $\sigma_l$, is available.
Usually this is done by fitting a Gaussian to a given spectrum \citep{2004PASP..116..138C,2011MNRAS.417.1787F,2014MNRAS.440.1634P}.
In this case, the dynamical mass within the effective radius $R_e$ is often simply estimated as:
\begin{equation}
M_\mathrm{dyn}=K \frac{R_e \sigma_l^2}{G}
\label{mdyn_k}
\end{equation}
where $G$ is the gravitational constant, $\sigma_l$ the los-velocity and $K$ a dimensionless parameter
\citep{2002A&A...386..149B,2001ApJ...554..981P,1991ApJ...376..380C,1998ApJ...508..539P,2006ApJ...646..107E,2006MNRAS.366.1126C,2009ApJ...704.1274W,2010MNRAS.406.1220W,2012MNRAS.423..632F,2014RvMP...86...47C,2014MNRAS.440.1634P,2016ApJ...831..149W,2017MNRAS.469.2335C}.
The value of $K$ is uncertain and different values have been used \citep{2002A&A...386..149B,2006MNRAS.366.1126C,2009ApJ...704.1274W,2010MNRAS.406.1220W,2017MNRAS.469.2335C}. A commonly used value is $K=5$, or a similar constant value \citep{1998ApJ...508..539P, 2001ApJ...554..981P,2006MNRAS.366.1126C,2009ApJ...704.1274W}, which in some cases lead to non-physical results where the estimated total mass was smaller than the stellar mass: $M_\mathrm{dyn}<M_\star$ \citep{2006ApJ...646..107E,2012MNRAS.423..632F,2014MNRAS.440.1634P}.

In this paper we use analytic arguments and high-resolution cosmological zoom-in simulations to estimate the parameters $\alpha$ and $K$, necessary for mass-estimates from observations where reliable values for rotation and dispersion are obtained and for observations where only a line-of-sight velocity dispersion is available.
Since all of these mass-estimates rely on the assumption of dynamical equilibrium we first have to validate Jeans equilibrium for the stars and the equivalent hydrostatic equilibrium for the gas. Then we measure for the simulated galaxies the velocities. Combining these with the measured total mass within the relevant radius, we obtain $\alpha$ and $K$ as a function of observables. We investigate the importance of additional terms including a gradient in $\sigma_r(r)$, a non-spherical potential and anisotropic velocity dispersion.
We do this for different structural components of the galaxies, namely the disc and the spheroid, gas and stars.

Galaxies are highly perturbed and gas rich at high redshifts. In many cases the disc is intensely fed by incoming intense streams \citep{2009Natur.457..451D}.
Through episodes of dissipative gas contractions, the gas can be quickly pushed towards the central region of the galaxies, leading to the formation of compact, gas-rich, star-forming blue nuggets \citep{2015MNRAS.450.2327Z,2016MNRAS.457.2790T,2016MNRAS.458..242T,2020MNRAS.496.5372D,2020MNRAS.493.4126D}. This phenomenon has been been referred to as wet compaction \citep{2014MNRAS.438.1870D}. Those objects are small in size but very massive, star forming systems.
Through intense star formation, the inner gas supply will be exhausted and the Blue Nuggets will gradually turn into quenched Red Nuggets with low star formation. Cosmic gas streams feed the galaxy at the same time with fresh cold gas. A galactic disc is slowly formed \citep{2016MNRAS.457.2790T,2016MNRAS.458..242T}.
These characteristic episodes from compaction events into Nuggets and finally disc like systems, surrounding the compact spheroids, are a robust phenomena that are accompanied by distinctive changes in morphology and kinematics. It has been shown that one very distinct feature is a clear critical stellar mass of $M_\star \sim 10^{9.5} \mathrm{M}_\odot$ for the major event of compaction followed by quenching.

Galactic gas discs are likely to survive only in dark-matter halos of mass above a threshold
of $\sim2 \times10^{11} M_\odot$, corresponding to a stellar mass of $\sim10^9 M_\odot$.

This is mostly because in the low-mass regime, the angular momentum is predicted to flip on a timescale shorter than the orbital timescale due to mergers \citep{2020MNRAS.493.4126D}. Additionally, violent disc instability exerts torques that shrink the disc by removing angular momentum. Lastly supernova feedback plays an important role in disrupting discs below the critical mass \citep{1986ApJ...303...39D}. 
Above the threshold mass, disruptive merger events are less frequent and not necessarily associated with a change in the pattern of the feeding streams. At the same time, the effects of supernova feedback are reduced \citep[e.g.][]{2020MNRAS.492.1385K,2020MNRAS.497.4346K}. All these effects allow gas discs to survive.
Other changes are the transition from diffuse to compact with an extended disc and envelope, from prolate to oblate, from pressure to rotation support, from low to high metallicity, and from supernova to AGN feedback \citep{2015MNRAS.450.2327Z,2019arXiv190408431D}.

We have structured this paper as follows:
First, in \autoref{equilibrium}, we describe the simulation setup and derive analytic expressions that we will use to investigate the assumptions for Jeans- and hydrostatic-equilibrium.
Secondly, in \autoref{sec:alpha}, we focus on the case when both rotation velocity and velocity dispersion ($v_\phi$ and $\sigma_r$) are measured to obtain $\alpha$. We focus on Jeans- and hydrostatic equilibrium in different components and we investigate if models for self-gravitating discs are valid and applicable.
In \autoref{sec:K} we focus on cases where only $\sigma_l$ is available to obtain $K$ both for stars and gas.
We finally summarise our results and conclude in \autoref{sec:conc}.

\section{Jeans equilibrium in the VELA simulations}
\label{equilibrium}
Our methodology is as follows:
In the first subsection, we describe the simulation setup that we will use in our analysis.
Then we recall the Jeans equation and derive analytic expressions for $\alpha$ for specific cases.
We describe how we obtain the quantities of interest from the simulation and how we use them to infer if Jeans or hydrostatic equilibrium is valid.
Finally, we derive three correction terms in the Jeans equation.

\subsection{Simulation Method and Subgrid Physics}
\smallskip
Alongside our analytical modelling, we use a series of VELA zoom-in hydro-cosmological simulations at $z=1-5$, which are described in Appendix \ref{sec:app_vela}, \autoref{tab:sample} and previous works \citep[e.g.][]{2014MNRAS.442.1545C,2015MNRAS.450.2327Z,2019MNRAS.488.4753D,2020MNRAS.493.4126D,2020MNRAS.496.5372D}.

The simulations were performed using the Adaptive Refinement Tree (\texttt{ART}) code \citep{1997ApJS..111...73K,2009ApJ...695..292C}.
The suite comprises 34 galaxies evolved to $z\!\sim\! 1$, with a maximum spatial resolution of $17.5$ to $35 \pc$ at any given time.
The dark matter halo masses at $z\!=\!2$ span from $10^{11}$ to $10^{12}\mathrm{M}_\odot$.
The choice of galaxies was made such that their dark matter halos have not undergone a major merger close to $z\!=\!1$.

Additionally, the code contains a set of subgrid physics models that describe many relevant processes of galaxy formation that are not directly calculable because of the limited resolution \citep{2009ApJ...695..292C,2012MNRAS.420.3490C,2015MNRAS.450.2327Z,2014MNRAS.443.3675M}. Those processes include gas cooling by atomic hydrogen and helium, metal and molecular hydrogen cooling, photoionization heating by the UV background with partial self-shielding, stochastic star formation, stellar feedback, metal enrichment, stellar mass loss, thermal feedback from supernovae, stellar winds, gas recycling and an implementation of feedback from radiation pressure as described in \cite{2014MNRAS.442.1545C}.

\subsection{Analytic expressions for \texorpdfstring{$\alpha$}{alpha}}
Consider a cylindrically symmetric galaxy that is rotating.
Assume that along the radial direction the stellar component is in Jeans equilibrium and the gas component is in hydrostatic equilibrium \citep{1915MNRAS..76...70J,2008MNRAS.383.1343W,2010ApJ...725.2324B,2010MNRAS.407.1148C,2020MNRAS.497.4051W}. This implies that at any given radius $r$, the gas and stars each obey
\begin{equation}
V_c^2(r) = v_\phi^2(r) - \frac{1}{\rho(r)} \frac{\diff (\rho(r) \sigma_r^2(r))}{\diff \ln r},
\label{eq:isotropic_jeans}
\end{equation}
where the circular velocity $V_\mathrm{c}$ represents the gravitational force through the potential gradient
\begin{equation}
V_c^2(r) = r \frac{\partial \Phi(r)}{\partial r} \backsimeq \frac{G M(<r)}{r}
\label{gm/r}
\end{equation}
and where the last expression is accurate for a spherical system, with $M(<r)$ the total mass within a sphere of radius $r$. The velocity $v_\phi(r)$ is the actual angular-averaged rotation speed in the disc plane (perpendicular to the angular-momentum vector), representing the centrifugal force.
The last term represents the pressure gradient in the disc, where $\rho(r)$ is the 3D density profile of the given component and $\sigma_r$ is the radial velocity dispersion of that component (with the thermal pressure assumed to be negligible for the gas).
We additionally assume that the velocity dispersion is isotropic. 
\autoref{eq:isotropic_jeans} is valid for the stars and the gas separately. 
\subsubsection{The Jeans Equation with Constant Dispersion}
Assuming the radial velocity dispersion $\sigma_r^2$ to be constant with radius, \autoref{eq:isotropic_jeans} can be written:
\begin{equation}
V_c^2 = v_\phi^2 + \alpha\, \sigma_r^2
\label{eq:alpha}
\end{equation}
with $-\alpha$ the logarithmic slope of the density profile,
\begin{equation}
\alpha \equiv - \frac{\diff \ln \rho}{\diff \ln r}.
\label{alpha_def}
\end{equation}
For example, for an isothermal sphere $\alpha = 2$, but different values for different 3D density profiles within the disc are obtained.
\subsubsection{Self-gravitating Exponential Disc}
For a self-gravitating disc, following \citet{2010ApJ...725.2324B}, if the velocity dispersion $\sigma$ is isotropic and constant also in the $z$ direction, the vertical density distribution $\rho(z)$ is given by the vertical hydrostatic Spitzer solution \citep[Section 4]{1942ApJ....95..329S,2008gady.book.....B}:
\begin{equation}
\rho(z)=\rho_0 \mathrm{sech}^2(z/h)
\end{equation}
where $\rho_0(r)$ is the density at the mid-plane $(z=0)$ and the scale height is
\begin{equation}
h(r)=\frac{\sigma}{\sqrt{2\pi G \rho_0(r)}}.
\end{equation}
The surface density $\Sigma(r)$ for such a scenario is given by
\begin{equation}
\Sigma(r) = 2 \rho_0(r) h(r).
\end{equation}
Combining the two Equations above yields for the density
\begin{equation}
\rho_0(r) = \frac{\pi G \Sigma^2(r)}{2\sigma^2}.
\end{equation}
After inserting in the definition of $\alpha$ (\autoref{alpha_def}), we get
\begin{equation}
\alpha(r) = - 2 \frac{\diff \ln \Sigma(r)}{\diff \ln r}.
\end{equation}
For the special case of an exponential disc with $\Sigma(r)=\Sigma_0 \exp(-r/r_d)$, where the half-mass radius is related to the exponential radius $r_d$ by $R_\mathrm{e} = 1.68 r_d$, one obtains
\begin{equation}
\alpha(r) = 2 \frac{r}{r_\mathrm{d}} = 3.36 \frac{r}{R_\mathrm{e}}.
\label{eq:burkert}
\end{equation}
The value of $\alpha$ at the effective radius is larger than that of an isothermal sphere, indicating that one should not expect the same value of $\alpha$ for all galaxy types or all components.
\subsubsection{Sérsic Profile Model}
\label{sec-sersic}
One can determine $\alpha$ for a Sérsic profile as a function of the Sérsic index $n$. Consider the Sérsic profile that describes the two-dimensional surface density $\Sigma(r)$ as a function of the two-dimensional radius $r$, with the parameter $n$, the Sérsic index describing the steepness of the profile \citep{1963BAAA....6...41S}. The profile is given by
\begin{equation}
\Sigma_s(r,n)=\Sigma_e \exp \left[ -b(n)\left[\left(\frac{r}{R_\mathrm{e}}\right)^{1/n}-1\right] \right],
\label{Sersic_formula}
\end{equation}
where $b(n)$ is chosen such that $R_\mathrm{e}$ is the half-mass radius. Therefore $\Sigma_e$ is the surface density at the half-mass radius. The function $b(n)$ can be approximated as \citep{2005PASA...22..118G,2015MNRAS.454.1000G}:
\begin{equation}
b(n) \simeq 1.9992n-0.3271.
\end{equation}
(See \cite{1999A&A...352..447C} for an alternative approximation.)

Choosing $n=1$ yields the surface density profile of an exponential disc. A de Vaucouleurs profile, characteristic of an elliptical galaxy, is obtained if $n=4$ \citep{1948AnAp...11..247D}. Integrating the two-dimensional Sérsic profile gives the three-dimensional density profile within the disc:
\begin{equation}
\rho(r',n) = - \frac{1}{\pi}\int_{r'}^{\infty}\frac{\diff\Sigma_s(r,n)}{\diff r}\frac{1}{(r^2-r'^2)^{1/2}}\diff r,
\end{equation}
where $r'$ is the spherical radius.
Here we have inverted the formula for the projection of the density to a surface density with Abel's formula \citep{1991ApJ...376..380C,2008gady.book.....B}.
From this $\alpha(n)$ can be obtained numerically as a function of Sérsic index $n$ with \autoref{alpha_def}:
\begin{equation}
\alpha(n) = - \frac{\diff \ln \rho(r,n)}{\diff \ln r}.
\label{alpha_n}
\end{equation}
This is shown in \autoref{fig:all_n_combo} for different radii.

\subsection{Testing Jeans/hydrostatic equilibrium: \texorpdfstring{$\alpha_v$}{alpha v} and \texorpdfstring{$\alpha_\rho$}{alpha rho}}
Rearranging \autoref{eq:alpha}, we first measure from the velocities
\begin{equation}
\alpha_\mathrm{v} = \frac{V_\mathrm{c}^2-v_\phi^2}{\sigma_r^2}.
\label{alpha_v_def}
\end{equation}
The rotational and radial velocities $v_\phi$ and $v_r$ are calculated using the coordinates $x,y$ in the plane of rotation
\begin{equation}
v_{\phi} = (x v_y - y v_x)/r, \quad v_{r} = (x v_x + y v_y)/r,
\end{equation}
with $r=\sqrt{x^2+y^2}$. The circular velocity $V_c$ is calculated with the total mass enclosed in a sphere. The radial velocity dispersion is calculated as $\sigma_r = \sqrt{\langle v_r^2\rangle-\langle v_r\rangle^2}$ where the average is computed in a cylindrical ring.
We note that if $\sigma_r$ is measured over segments of cylindrical rings, a smaller value is obtained. However, it has been shown \citep[see section 4.2 in][]{2020MNRAS.497.4051W} that such an estimate under-predicts the pressure term in the Jeans equation.
In comparison, using \autoref{alpha_def}, we measure from the density profile
\begin{equation}
\alpha_\rho = - \frac{\diff \ln \rho}{\diff \ln r},
\end{equation}
where the density $\rho(r) = (4\pi r^2)^{-1} \diff M(r)/\diff r$ is calculated from the smoothed cumulative mass profile $M(r)$.
If Jeans or hydrostatic equilibrium is valid and the velocity dispersion is constant, \autoref{eq:alpha} is valid and $\alpha_\mathrm{v}$ and $\alpha_\rho$ are identical. Therefore we can validate the assumption of equilibrium based on the agreement between $\alpha_\mathrm{v}$ and $\alpha_\rho$.

\subsection{Correction Terms in the Jeans Equation}
Deviations may arise if one of the above stated assumptions are not valid.
We therefore include the following corrections.
\subsubsection{Non-Constant Velocity Dispersion}
First, we include the term
\begin{equation}
\gamma \equiv -\frac{\diff \ln \sigma_r^2}{\diff \ln r}
\end{equation}
to account for non-constant velocity dispersion in \autoref{eq:isotropic_jeans}.
\subsubsection{Non-Spherical Potential}
The above given expression $V_c^2 \simeq GM/r$ is only true for a spherical mass distribution.
To allow for a more accurate expression we expand the potential $\Phi$ in multipoles to find non-spherical correction terms.
We write the potential at position ${\bf r}$ as
\begin{equation}
\Phi(\mathbf{r}) = - \frac{G}{r} M - \frac{G}{r^3} \mathbf{r}_\alpha D_\alpha - \frac{G}{2 r^5} \mathbf{r}_\alpha \mathbf{r}_\beta Q_{\alpha \beta} - \ldots
\label{quadropole}
\end{equation}
where $r=|\mathbf{r}|$ and summation is implied over double occurring coordinates $\alpha, \beta = {x,y,z}$. The terms $M, D_\alpha$ and $Q_{\alpha \beta}$ are the monopole, the dipole and the (traceless) quadrupole:
\begin{equation}
\begin{split}
M &= \sum_i m_i,\\ 
D_\alpha &= \sum_i m_i \mathbf{r}_{i,\alpha},\\
Q_{\alpha \beta} &= \sum_i m_i(3 \mathbf{r}_{i,\alpha} \mathbf{r}_{i,\beta}-\delta_{\alpha \beta} r_i^2).
\end{split}
\end{equation}
Since the monopole represents the total mass and the dipole vanishes, the quadrupole is the first term that contributes as a non-spherical correction to the potential. 

To obtain the quadrupole correction in the symmetry plane of the disc at the effective radius $R_\mathrm{e}$,
we calculate the eigenvalues of $Q_{ij}$.
The correct eigenvalues are identified using the properties that the quadrupole is by construction trace-less and the trace is invariant under rotation.
Including contributions from the non-spherical part of the potential, the circular velocity $V_{c,Q}$ is changed and therefore also the resulting $\alpha_{\mathrm{v},Q}$. We define the difference as 
\begin{equation}
\Delta_Q\equiv\alpha_\mathrm{v}-\alpha_{\mathrm{v},Q}.
\end{equation}
\subsubsection{Anisotropic Velocity Distribution}
Lastly, we include corrections for anisotropic dispersion where we define the anisotropy parameter as
\begin{equation}
 \beta \equiv 1 - \frac{\sigma_\phi^2 }{ \sigma_r^2}.
\end{equation}
Note that $\beta$ as defined here is actually twice the standard $\beta$ in the cylindrical case \citep[see Eq. 4.61 and 4.224 in][]{2008gady.book.....B}.

\begin{figure}
\center
\includegraphics[width=\columnwidth]{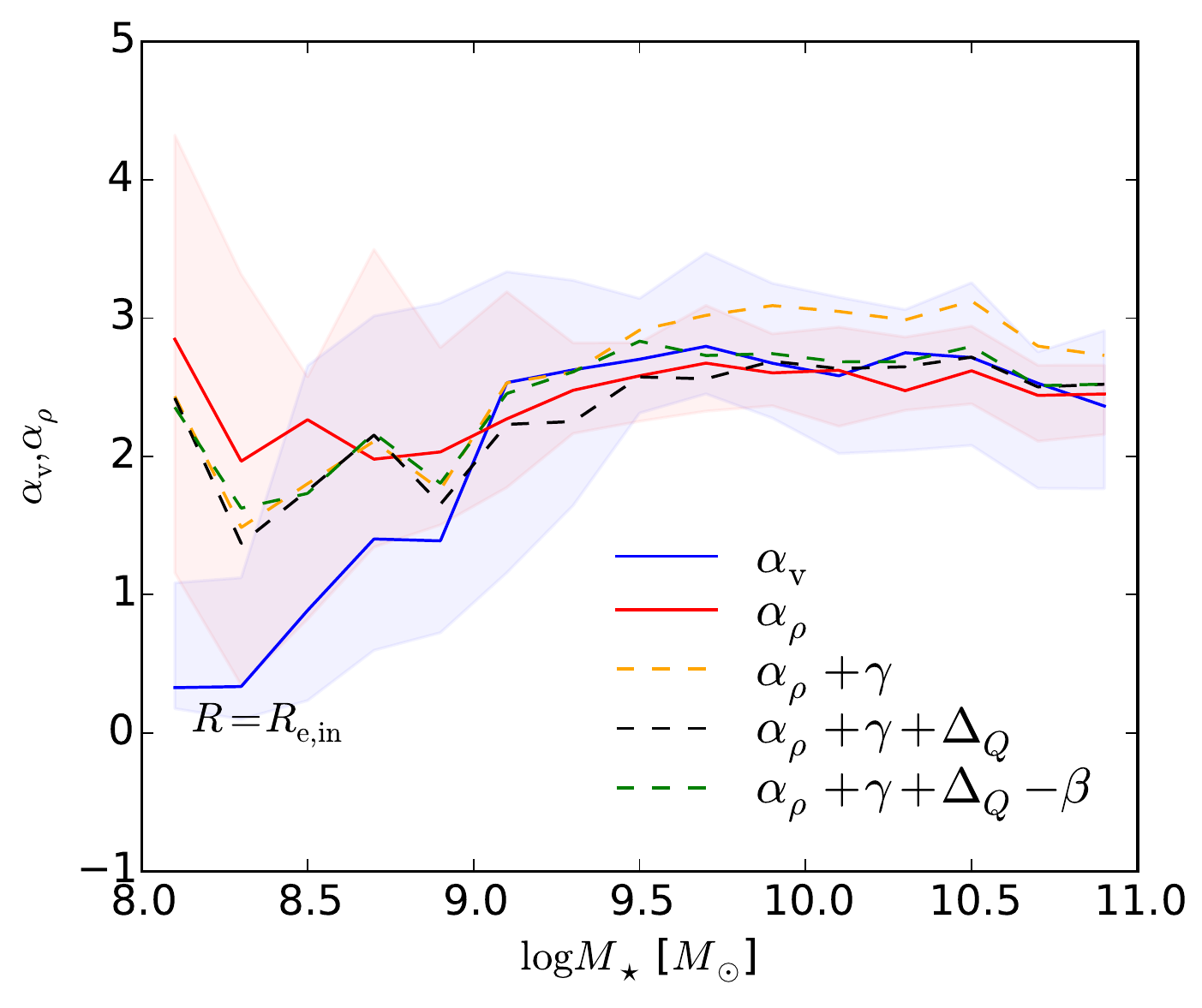}
\caption{
Validity of the Jeans equation for the bulge as a function of stellar mass.
Values for $\alpha_\rho$ obtained from the density profile and $\alpha_\mathrm{v}$ obtained from the velocities for the galactic bulge at the effective radius $R_\mathrm{e,in}$ with additional corrections. Shown is the median of the obtained values and the $68\%$ confidence level for $\alpha_\rho$ and $\alpha_\mathrm{v}$.
Curves agree after compaction ($M_\star > 10^{9.5} \mathrm{M}_\odot$) which indicates Jeans equilibrium with $\alpha \sim 2.55$. The corrections for a non-constant dispersion $\gamma$, non-spherical potential $\Delta_Q$ and anisotropic dispersion $\beta$ are included.
The Jeans equation is valid for $M_\star > 10^{9.5} \mathrm{M}_\odot$.}
\label{fig:alpha_vs_m_re_in}
\end{figure}

\smallskip
Taking the three corrections mentioned above into account, \autoref{eq:isotropic_jeans} becomes
\begin{equation}
V_c^2=v_\phi^2 + \sigma_r^2(\alpha_\rho + \gamma + \Delta_Q - \beta).
\label{all_corr_incl}
\end{equation}
From this, together with the definition of $\alpha_\mathrm{v}$ (\autoref{alpha_v_def}), we obtain
\begin{equation}
\alpha_\mathrm{v} = \alpha_\rho + \gamma + \Delta_Q - \beta.
\label{rearr}
\end{equation}
By comparing the two sides of this equation (see for example \autoref{fig:alpha_vs_m_re_in}) we can evaluate the degree of validity of the Jeans equation or the equation for hydrostatic equilibrium respectively, and the relative contribution of each correction. If equilibrium is approximately valid, we can estimate $M_\mathrm{dyn}$ by using the values for $\alpha_\mathrm{v}$ in \autoref{all_corr_incl},
\begin{equation}
V_c^2 = v_\phi^2 + \alpha_\mathrm{v}\, \sigma_r^2.
\end{equation}

\section{Measuring the dynamical mass from decomposed kinematics}
\label{sec:alpha}
\begin{figure*}
  \includegraphics[width=.7\textwidth]{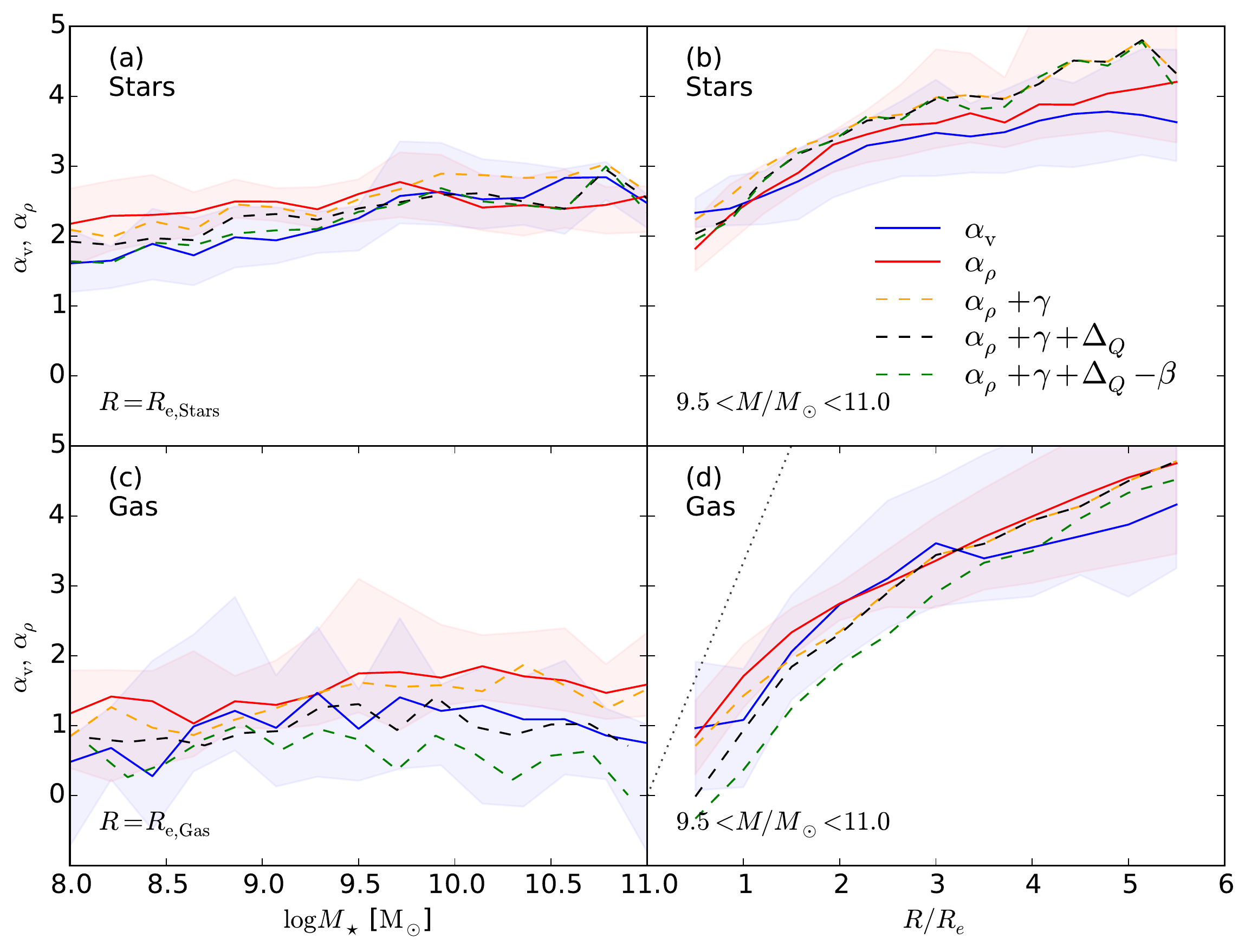}
  \caption{Validity of the Jeans/hydrostatic equilibrium in the disc.
  Values for $\alpha_\rho$ obtained from the density slope and $\alpha_\mathrm{v}$ obtained from the velocities in the galactic disc.
  (\textbf{a}) The evolution of measured values for the stellar system as a function of stellar mass at $R_\mathrm{e,stars}$.
  The agreement of the lines indicates Jeans equilibrium.
  (\textbf{b}) The radial profile of the stars above $10^{9.5} \mathrm{M}_\odot$ (post compaction) indicates Jeans equilibrium up to $\sim~5R_\mathrm{e,stars}$.
  (\textbf{c}) The evolution of measured values for the gas as a function of stellar mass at $R_\mathrm{e,gas}$. The agreement of the lines indicates crude hydrostatic-equilibrium.
  (\textbf{d}) The radial profiles of the gas above $10^{9.5} \mathrm{M}_\odot$ (post compaction), indicating hydrostatic equilibrium out to $3.5R_\mathrm{e,gas}$, and crude equilibrium out to $\sim 5 R_\mathrm{e,gas}$.
The profile deviates significantly from the prediction for an self-gravitating exponential disc which is shown as grey dotted line (see Figs. \ref{fig:accel-disc}, \ref{fig:accel-vs-alpha}).}
  \label{fig:alphadisc}
\end{figure*}

We now report on our findings for equilibrium in different components of the galaxies. First we focus on bulges, secondly we investigate equilibrium in galactic discs and lastly we analyse equilibrium in whole galaxies.
Using the VELA suite of high resolution cosmological simulations we test if stars and gas in high-redshift galaxies are in Jeans or hydrostatic equilibrium.
We compare values for $\alpha_\mathrm{v}$ obtained from the velocities and $\alpha_\rho$ obtained from the density slope. As described above, equilibrium is indicated if those values agree.
The analysis is done in steps of scale factor $\Delta a = 0.01$ using the entire VELA suite in the redshift range $z=1-5$, which includes $933$ snapshots. 
We do not show the evolution of the measured quantities with redshift explicitly, but report instead on the evolution with stellar mass, since the two quantities are strongly correlated. Furthermore, the stellar mass provides an important indication of the evolutionary phase of galaxies, since below a stellar mass of $10^{9.5} \mathrm{M}_\odot$ galaxies are highly perturbed and only above this critical mass long lived discs are expected.

\subsection{Jeans Equilibrium in Galactic Bulges}
\label{sec:bulge}

First, we are interested in verifying the assumption of Jeans equilibrium for stars in bulges of galaxies.
We focus on Jeans equilibrium of the stars only, since there is little gas left in the central part of galaxies.
To separate the galactic bulge from the disc we fit a double Sérsic profile to the surface density profile of the stars face-on, where the Sérsic index for the outer disc part is kept constant at $n_\mathrm{out}=1$ (see \autoref{sec:sersic}).
Although stars from the bulge dominate at $R_\mathrm{e,in}$, there are also stars belonging to the disc present. However, disc stars (as defined in \autoref{sub:self-grav}) inside $R_\mathrm{e,in}$ on average only make up $8.5\%$ of the total bulge mass and therefore will not significantly change our results.

At the resulting effective radius $R_\mathrm{e,in}$ of the bulge, the velocities are measured and mass weighted averaged in rings centred on $R_\mathrm{e,in}$ with size $\Delta R = \pm R_\mathrm{e,in} /2$. The plane of rotation is determined by the angular momentum inside $R_\mathrm{e,in}$. 
To verify if the assumption of Jeans-equilibrium is valid, we show the values for $\alpha_\mathrm{v}$ obtained from the velocities and $\alpha_\rho$ obtained from the density slope in \autoref{fig:alpha_vs_m_re_in}.
For $M_\star>10^{9.5}M_\odot$ the similarity between $\alpha_\mathrm{v}$ and $\alpha_\rho$ indicate the validity of Jeans equilibrium with a scatter of $\sim 20\%$ around a value of
\begin{equation}
\alpha_\mathrm{v}=2.63\pm0.53,
\end{equation}
constant in stellar mass.
This is consistent with the finding that long-term discs form in these simulations above a threshold mass of $M_{\rm vir} \sim 10^{11}\msun$, due to the infrequent merger-driven spin flips and compaction-driven bulges \citep{2020MNRAS.493.4126D}.

Including $\gamma$ as a correction for a non-constant dispersion does not significantly improve the validity of Jeans equilibrium at low masses, and generates a slight overestimate by $\sim 0.4$ of $\alpha_\rho$ at high masses ($\gamma=0.38\pm 0.25$). This discrepancy is alleviated if the correction for a non-spherical potential are included ($\Delta_Q = -0.31 \pm 0.18$). This is because a deeper potential well will be counteracted by a larger dispersion. Accounting for anisotropic dispersion through $\beta$ does not change the results significantly ($\beta=-0.01 \pm 0.16$).

In galaxies with stellar masses below $10^{9.5}\mathrm{M_\odot}$, before compaction and where discs are disrupted by frequent merger-driven spin flips, $\alpha_\mathrm{v}$ systematically underestimates $\alpha_\rho$, by an amount that is comparable to the 1-$\sigma$ scatter among galaxies.
The corrections for a non-spherical potential or anisotropic dispersion and non-constant dispersion are not significant ($\Delta_Q = -0.05 \pm 0.13$, $\beta=-0.03 \pm 0.20$, $\gamma = -0.10\pm 0.66$).
Below $M_\star=10^{9.5}M_\odot$, the value for $\alpha$ can be crudely approximated with 
\begin{equation}
\alpha_\mathrm{v}(M_\star)= 1.8 \log( M_\star / 10^{9.5} \mathrm{M_\odot}) +2.63.
\end{equation}
For galaxies with stellar masses $M_\star<10^{8.5} \mathrm{M}_\odot$ the discrepancy between the curves is large and therefore the assumption of Jeans equilibrium is invalid.

\subsection{Equilibrium in the Plane of Galactic Discs}
\label{sec:disc}
\begin{figure}
  \includegraphics[width=\columnwidth]{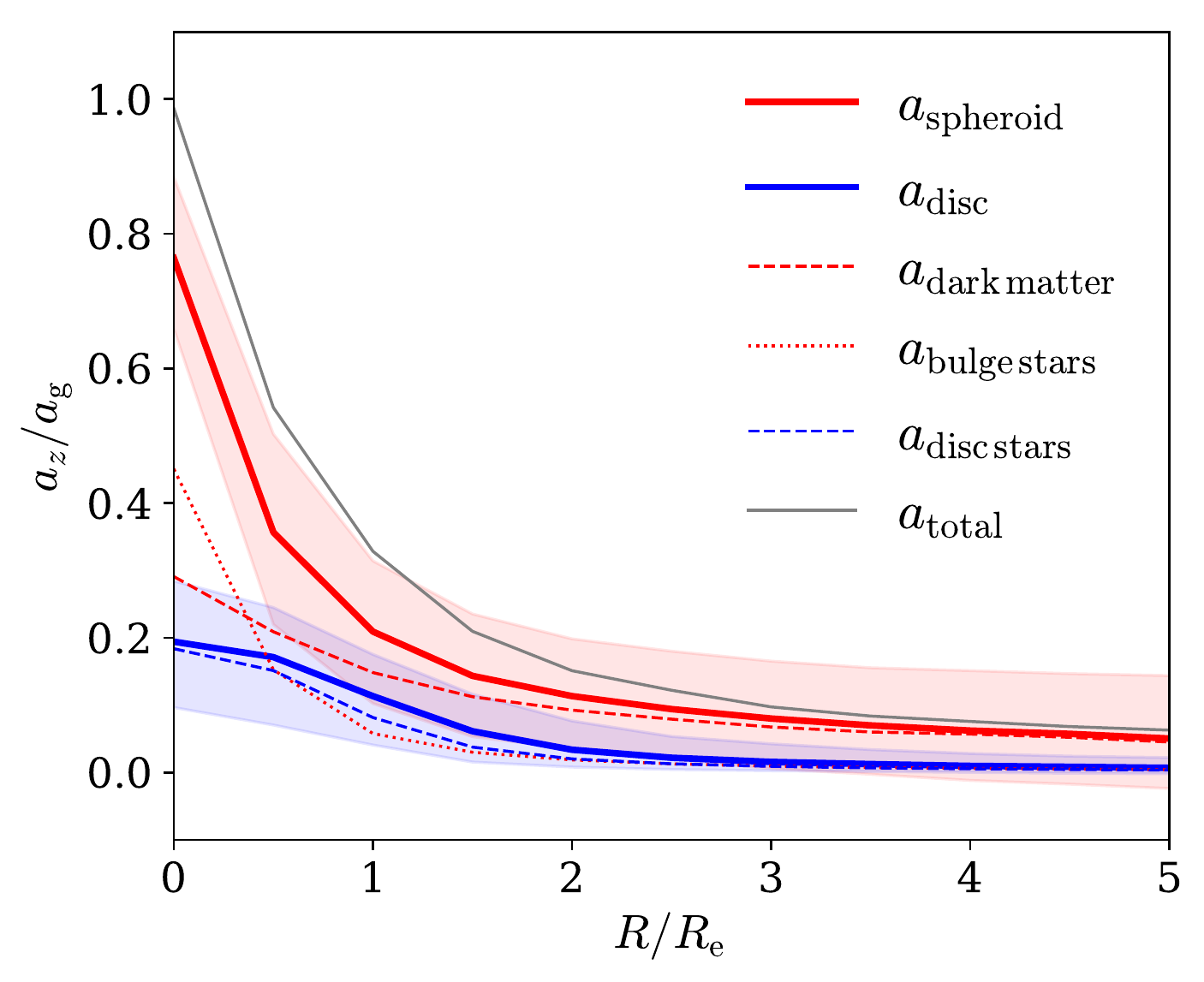}
  \caption{Is the approximation of self-gravitating disc valid? Shown is the specific force in the $z$ direction, $a_z$, in the plane $z=-H_d$, as a function of $R/R_e$, as exerted by different components of the galaxy (see labels). The force is normalised by $a_g=V_c^2(r)/r$. The half mass radius refers to the gas. The sample is limited to galaxies with $M_\star>10^{9.5}M_\odot$. We see that the contribution of the spheroidal component (dark matter plus bulge stars) dominates over the disc component (mostly stars) by a factor larger than two at $R<R_e$, and by a larger factor at large radii, implying that the approximation of self-gravitating disc is problematic.
  }
  \label{fig:accel-disc}
\end{figure}

\begin{figure}
  \includegraphics[width=\columnwidth]{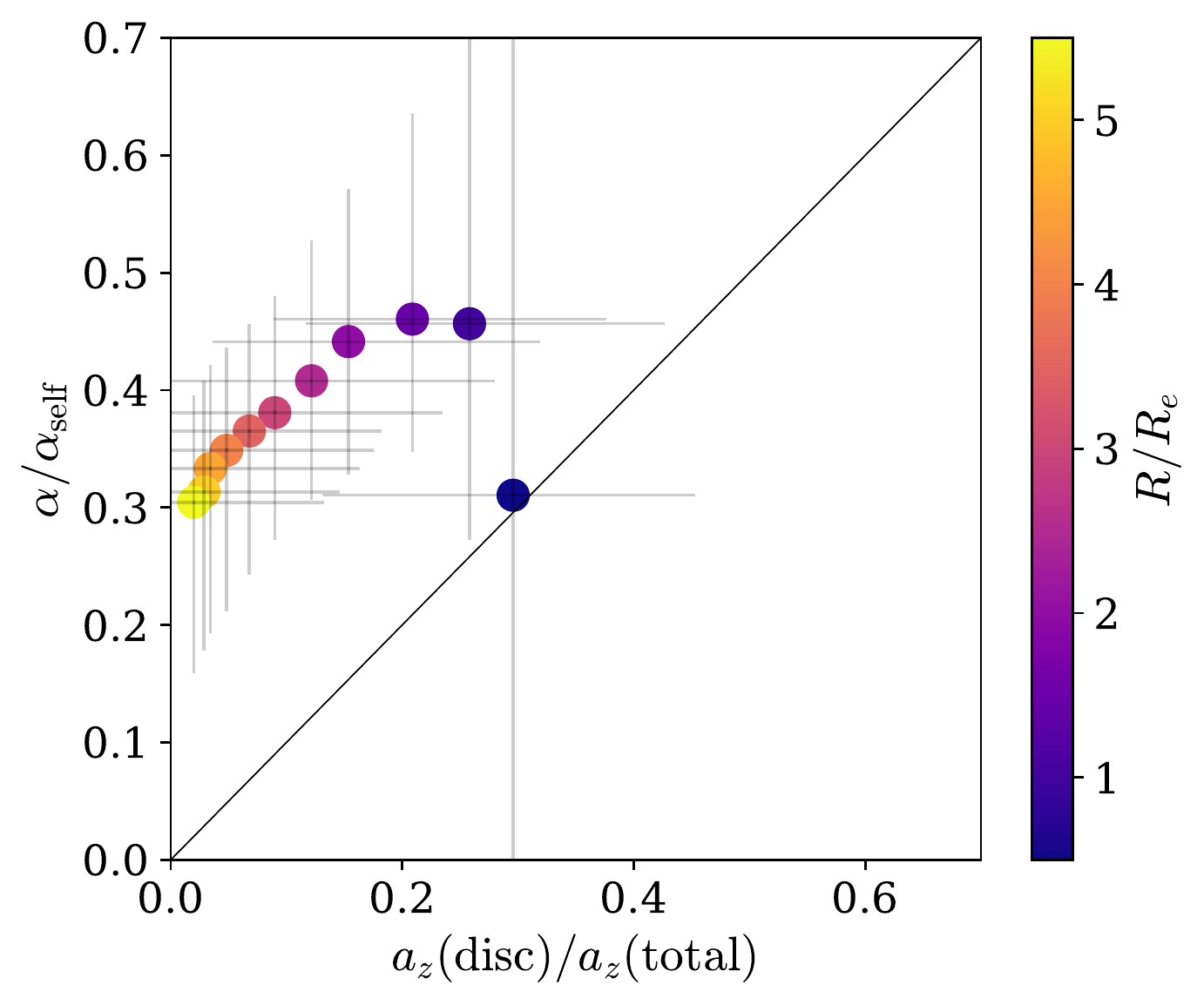}
  \caption{Validity of the self-gravitating disc approximation.
  Values for $\alpha_\rho$ from the simulations, divided by the prediction $\alpha_\mathrm{self}=3.36 (R/R_e)$ for self-gravitating discs, versus the specific force in the $z$ direction $a_z$ exerted by the disc with respect to the total $a_z$.
  Colours indicate radial bins, where the half mass radius refers to the gas.
  We learn that at $1-2 R_\mathrm{e}$ $\alpha$ is about $0.45 \alpha_\mathrm{self}$ (consistent with \autoref{fig:alphadisc}\hyperref[fig:alphadisc]{d}).
  }
  \label{fig:accel-vs-alpha}
\end{figure}

We expand our study of equilibrium to galactic discs. We are especially interested in the validity of Jeans and hydrostatic equilibrium in the plane of the disc. 
Therefore we choose a small height of $h=\pm R_e /4$ and use the 3D half-mass radius $R_\mathrm{e}$ of the analysed component (computed inside $0.1 R_\mathrm{vir}$).
The discs are orientated relative to the angular momentum of the analysed component, stars or cold gas ($T<5\times 10^4$K).
The velocities are mass weighted averaged over rings of width $0.5 R_e$ from $0.5 R_e$ to $5.5 R_e$.

\autoref{fig:alphadisc}\hyperref[fig:alphadisc]{a}
shows the values of $\alpha_\rho$ and $\alpha_\mathrm{v}$ for the stars as a function of stellar mass at $R_\mathrm{e,stars}$. For the stars there is good agreement between $\alpha_\rho$ and $\alpha_\mathrm{v}$ for $M_\star>10^{9.5}M_\odot$, which indicates Jeans equilibrium with $\alpha \simeq 2.6$, constant for galaxies with large stellar masses. The small but systematic deviations for low mass systems indicate that deviations from Jeans equilibrium are not too large and the presented approach to $\alpha$ is still usable for crudely estimating the total mass. These discrepancies are alleviated if corrections are included.
For galaxies above the threshold mass the contribution from non-spherical potentials $\Delta_Q=-0.3 \pm 0.1$ is counteracted by the non-constant dispersion term $\gamma = 0.4 \pm 0.2$. Anisotropic dispersions are not significant ($\beta = 0.0 \pm 0.2$).

\autoref{fig:alphadisc}\hyperref[fig:alphadisc]{c} refers to the gas, where we show values of $\alpha_\rho$ and $\alpha_\mathrm{v}$ as a function of stellar mass at $R_\mathrm{e,gas}$.
We see large fluctuations of the radial velocity in rings, possibly reflecting variations along the rings and between the rings. 
However the average value for the radial velocity is consistent with zero ($v_r = -10.7 \pm 11.7\, \mathrm{km/s}$). (Note that in contrast, for the stars we find $v_r =-0.8 \pm 2.0\, \mathrm{km/s}$). These errors propagate to $\alpha_\mathrm{v}$ and the large size of the errors may indicate that there are deviations from equilibrium. We see that the median values for $\alpha_\mathrm{v}$ agree with $\alpha_\rho$, especially if non-spherical potentials and non-constant dispersions are taken into account ($\Delta_Q=-0.47 \pm 0.28$, $\gamma = -0.20 \pm 0.26$). We find that the velocity dispersion for the gas is not isotropic ($\beta=0.52 \pm 0.14$) and should be taken into account \citep{2009MNRAS.392..294A,2020A&A...639A.145C}.

For massive galaxies, we show radial profiles in \autoref{fig:alphadisc}\hyperref[fig:alphadisc]{b} for the stars and \autoref{fig:alphadisc}\hyperref[fig:alphadisc]{d} for the gas. For the stars it is apparent that there is equilibrium up to radii of $\sim 5 R_\mathrm{e,stars}$ where $\alpha$ increases from $\simeq 2.5$ to $\simeq 3.5$.
For the gas, the agreement between the curves imply that the assumptions of hydrostatic equilibrium are valid in the plane of the disc for radii up to $\sim 5 R_\mathrm{e,gas}$, where $\alpha$ increases from $\simeq 1$ to $\simeq 4$. The large errors however may indicate certain deviations from equilibrium. Increasing values of $\alpha$ with radius from $\simeq 1$ to $\simeq 4$ were also obtained by \cite{2020MNRAS.497.4051W} where four massive galaxies of the FIRE-2 suite were analysed. These findings imply that a radial gradient of the pressure is significantly contributing. We see that the radial gradient of the velocity dispersion $\gamma$ is insignificant in comparison to the slope of the density profile $\alpha_\rho$. This demonstrates that pressure support is mostly provided through the slope of the density profile.

There are large deviations from the predictions for an exponential disc (\autoref{eq:burkert}) where $\alpha$ was supposed to grow linearly with radius and should be e.g. $3.36$ at $R_\mathrm{e,gas}$ or $6.72$ at $2R_\mathrm{e,gas}$.
These deviations arise because the assumption of a self-gravitating disc is not valid, being embedded in a dark-matter halo (see below).

\subsection{Self-Gravitating Discs}
\label{sub:self-grav}
As discussed above, we have seen that values obtained for $\alpha$ in the disc deviate from the predictions for an exponential disc (\autoref{eq:burkert}).
To investigate this we first measure $a_z$, the specific force exerted by each component in the $z$ direction at a distance $z=-H_d$ from the disc plane, where $H_d$ is half the gas-disc height, as defined in \citet{2014MNRAS.443.3675M}. The specific forces are calculated by direct summation for test-points lying in the plane of $z=-H_d$ at various radii out to $5R_\mathrm{e,gas}$. We separately compute the contributions from dark-matter, stars and gas. Furthermore we divide stars into disc stars and spheroid stars, where for the disc we adopt a threshold of $J_z/J_c > 0.7$ for each star particle where $J_z$ is the angular momentum component parallel to the z-axis and $J_c$ the angular momentum of a co-rotating circular orbit with the same energy.

\autoref{fig:accel-disc} shows the specific force exerted by different components $a_z$ in the $z$ direction as a function of $R/R_\mathrm{e,gas}$ for galaxies with $M_\star > 10^{9.5} \mathrm{M_\odot}$ where gas discs are likely to survive.
When comparing contribution of the spheroidal components (bulge stars and dark matter) to the contribution of the disc components (gas and disc stars), we see that the spheroidal components dominates over the disc component by a factor of larger than two at $R<R_\mathrm{e,gas}$.
For $R>R_\mathrm{e,gas}$ this factor increases linearly in radius from $\sim 2$ at $R_\mathrm{e,gas}$ to $\sim 6$ at $5R_\mathrm{e,gas}$.
It is apparent that the gas disc contributes only little.
This is consistent with the $\alpha$ values that we obtain, which are very different from the self-gravitating disc prediction.
We see that the dark matter is the dominating component at radii $r>0.5 R_\mathrm{e,gas}$. Only inside $0.5 R_\mathrm{e,gas}$ the contribution from bulge stars becomes more dominant.

\autoref{fig:accel-vs-alpha} shows the obtained $\alpha$ from the simulations, divided by the prediction $\alpha_\mathrm{self}=3.36 (R/R_e)$ for self-gravitating disc given by \autoref{eq:burkert}.
It is shown versus $a_z$ by the disc with respect to the total $a_z$, with the colours indicating $R/R_\mathrm{e,gas}$.
It is apparent that near $R_\mathrm{e,gas}$ the relative contribution of the disc to $a_z$ is $\sim0.25$ and $\alpha \sim0.45\alpha_\mathrm{self}$ accordingly.
Namely $\alpha_\mathrm{self}$ under-estimates the contribution of the pressure term by a factor of 2.

At larger radii, the relative contribution of the disc drops even further, and $\alpha \sim 0.3\alpha_\mathrm{self}$ at $R\sim 5 R_\mathrm{e,gas}$.
The pressure term grows slower than linearly with radius.
We conclude that the approximation derived from a self-gravitating disc for the asymmetric-drift pressure term in the Jeans equation is not valid in our simulated high-$z$ discs.

\subsection{Jeans and Hydrostatic Equilibrium in Whole Galaxies}
\label{sec:alpha_all}
\begin{figure}
  \includegraphics[width=\columnwidth]{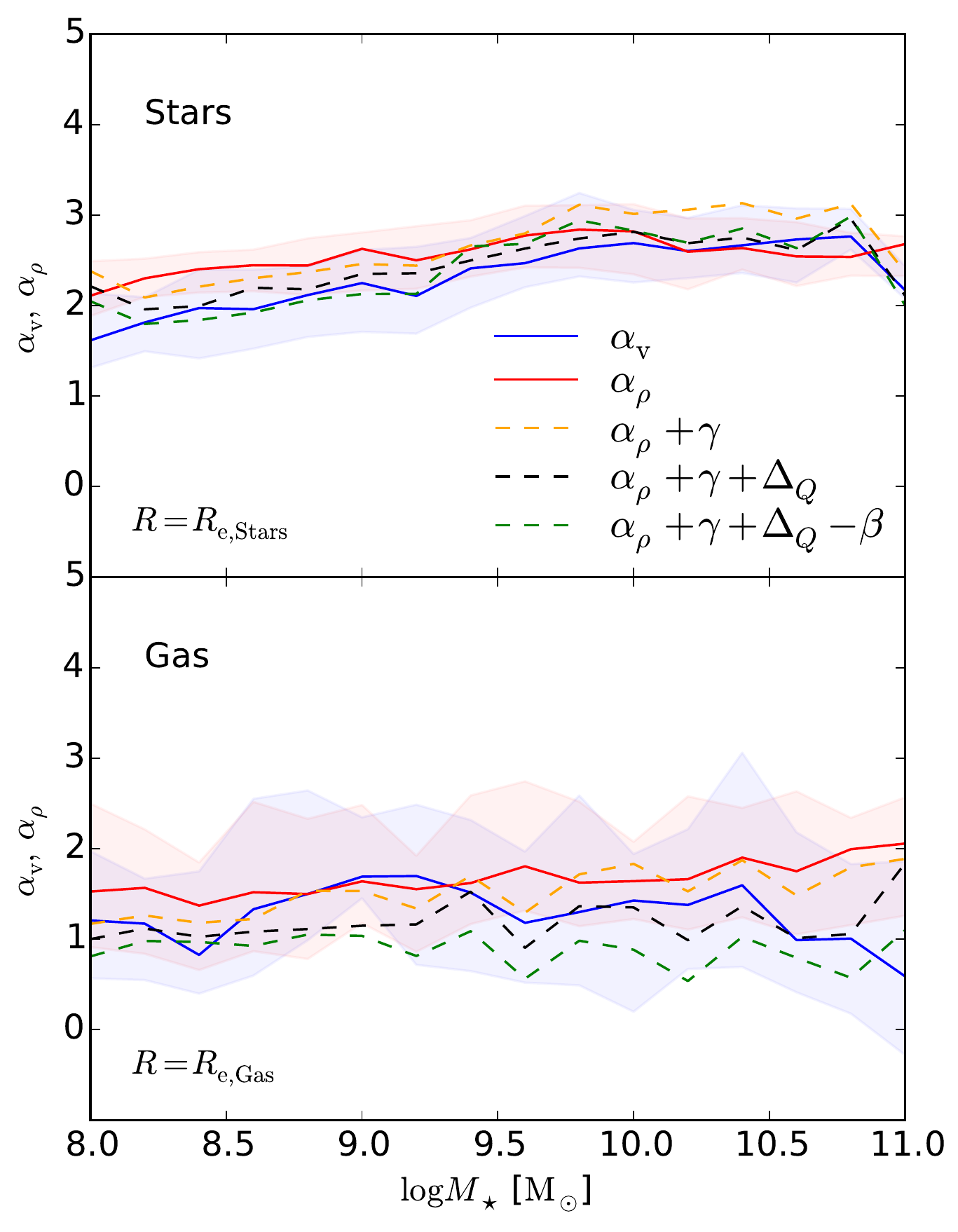}
  \caption{Values for $\alpha_\rho$ and $\alpha_\mathrm{v}$ for stars (top) and gas (bottom) evaluated at $R_e$ of the stars and the gas respectively. Here all the material inside a sphere is used - including stars/gas from the bulge and the disc - where in previous sections we separately analysed the bulge and the disc. For the stars, $\alpha_\mathrm{v}$ increases with mass from $\sim2$ to $\sim2.6$ at $M_\star < 10^{9.5}\mathrm{M_\odot}$ and is constant for more massive galaxies with $\alpha_\mathrm{v}\sim 2.6$.
  For the gas, at $M_\star < 10^{9.5}\mathrm{M_\odot}$ (pre-compaction), $\alpha_\mathrm{v} \sim 1.4$ and it decreases from $\sim 1.5$ to $\sim 0.5$ for more massive systems.
  Hydrostatic equilibrium at high masses is achieved only if the corrections are included.
  }
  \label{fig:both_all_in_re}
\end{figure}
Up until now, we only analysed the validity of equilibrium for stars and gas in the bulge and in the plane of the galactic disc.
Often it is however not possible to obtain kinematic measurements for gas or stars separately in the bulge or only in the thin plane of the disc. 
Therefore we now expand our study to galactic systems, namely where we use all the stars or gas inside a sphere with radius $R_e$, the half-mass radius of the analysed component.
The resulting values for $\alpha$ for stars and gas are shown in \autoref{fig:both_all_in_re}.

For the stars, we find for $M_\star > 10^{9.5}\mathrm{M_\odot}$, where there are long-lived discs,
\begin{equation}
\alpha_\mathrm{v}=2.61\pm0.35,
\end{equation}
in-line with the values that we obtained for the bulge only and for the disc only.
The correction for non-constant dispersion $\gamma=0.30\pm0.27$ is partly counteracted by the correction for non-spherical potential $\Delta_Q = -0.26\pm0.16$ and the correction for anisotropy $\beta=-0.04\pm0.15$ is negligible.
For galaxies below the threshold mass $(M_\star < 10^{9.5}\mathrm{M_\odot})$, $\alpha_\mathrm{v}$ increases with $M_\star$ from $\sim1.7$ to $\sim2.25$, $\gamma=-0.1\pm0.3$ is negligible but $\beta=0.14\pm0.16$ together with $\Delta_Q = - 0.12\pm0.07$ alleviate discrepancies with $\alpha_\rho$.
If all corrections are applied, we find Jeans equilibrium over the whole analysed mass-range. Polynomial approximations to the values of $\alpha$ can be found in \autoref{tab:alpha}.
These findings are consistent with results from a different suite of simulations (NIHAO), where the anisotropy $\beta$ was close to zero except very close to the centre and towards the virial radius such that anisotropy is negligible when assessing the dynamical mass and that the Jeans equation was valid with a $\sim13\%$ RMS error between 0.02 $R_\mathrm{vir}$ and 0.56 $R_\mathrm{vir}$ \citep{2020MNRAS.491.4523F}.

For the gas, $\alpha_\mathrm{v} \simeq 1.2$ for massive galaxies with $M_\star > 10^{9.5}\mathrm{M_\odot}$ (see the approximation in \autoref{tab:alpha}). There is a discrepancy with $\alpha_\rho$ which is alleviated if corrected for a non-constant dispersion $\gamma=-0.18\pm0.48$, and for non-spherical potentials $(\Delta_Q)(M_\star)= -0.307 \log( M_\star / 10^{9.5} \mathrm{M_\odot}) -0.228$ which evolves with mass. Including corrections for anisotropy $\beta=0.48 \pm 0.16$ leads to slightly lower predictions of $\alpha$. 
Below the threshold mass, the typical value for $\alpha_\mathrm{v} \simeq 1.4$ is in good agreement with $\alpha_\rho$ and the corrections are smaller ($\gamma=-0.07\pm0.9$, $\Delta_Q = - 0.18\pm0.09$, $\beta=0.27\pm0.31$).
\subsection{Sérsic profile}
\label{sec:sersic}
\begin{figure*}
  \includegraphics[width=0.65\textwidth]{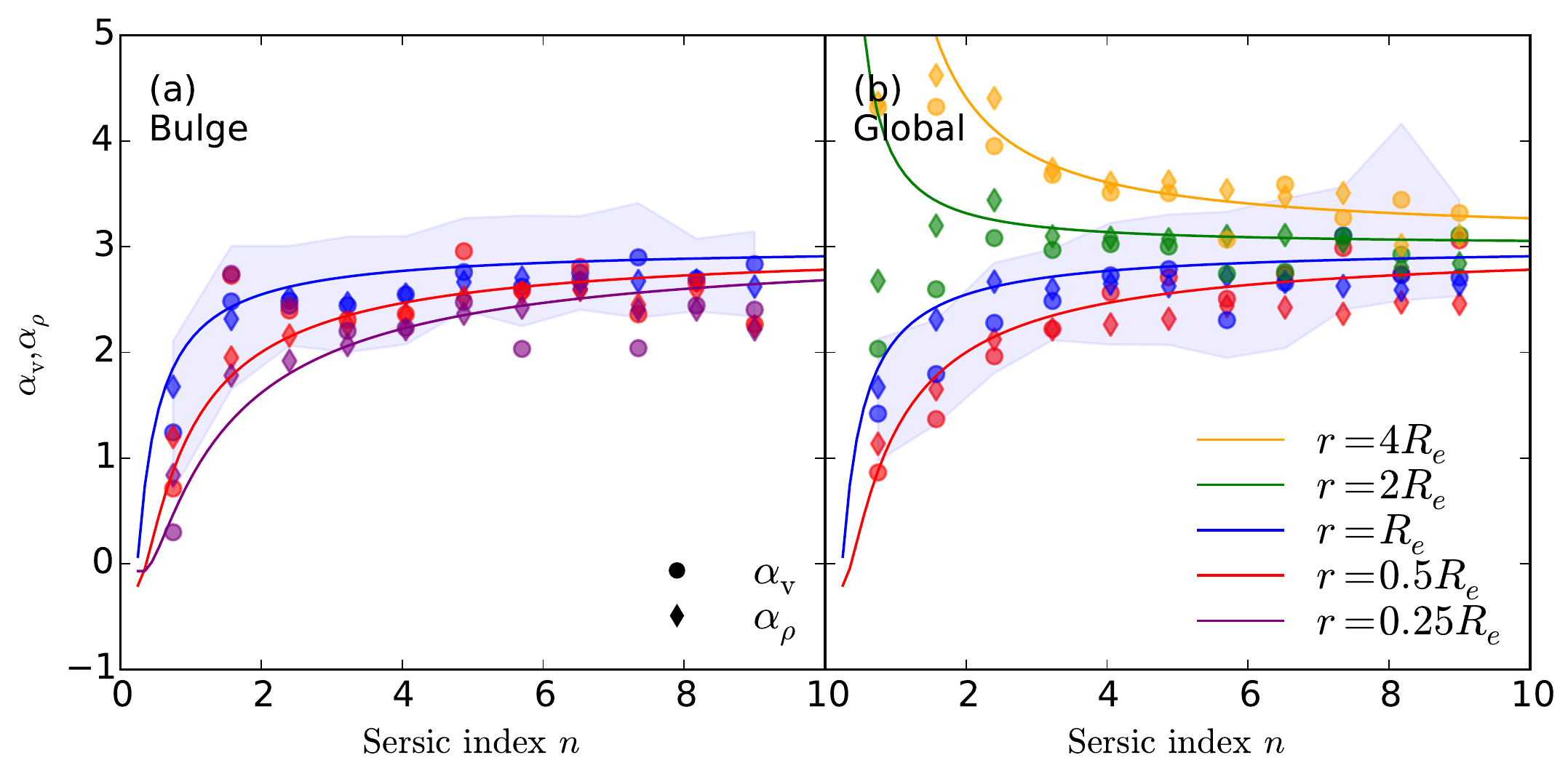}
  \caption{Dependence of $\alpha$ on Sérsic index.
  Values for $\alpha_\mathrm{v}$ and $\alpha_\rho$ for the stars shown as a function of Sérsic index $n$. Solid lines show theoretical expectations obtained by \autoref{alpha_n} for different radii. Measured values are shown as circles for $\alpha_\mathrm{v}$ and diamonds for $\alpha_\rho$. The shaded area indicates the size of the errors on $\alpha_\mathrm{v}$ as example for $r=R_e$. (\textbf{a}) The left panel shows results for the bulge. (\textbf{b}) The right panel shows results for whole galaxies.
  The half-mass radius refers to $R_\mathrm{e,in}$ for the bulge and $R_\mathrm{e,stars}$ for whole galaxies, obtained from the Sérsic fits.
  Measured values agree well with the theoretical expectations except for small $n$ at $r\geq2R_\mathrm{e}$.
  }
  \label{fig:all_n_combo}
\end{figure*}
In addition to $\alpha$ as a function of stellar mass we provide $\alpha$ as a function of Sérsic index. \autoref{fig:all_n_combo} shows the theoretical curves that we have derived in section \ref{sec-sersic}.
For an exponential disc, $\alpha(n=1)=2$ at the effective radius $R_\mathrm{e}$. For larger Sérsic indices, namely around $n=4$, which reflect a de Vaucouleurs profile, $\alpha(n) \simeq 2.7$ at $R_\mathrm{e}$.
For a fixed Sérsic index, $\alpha$ gets larger with increasing radius.
For $n>0.5$ the theoretical curves can be approximated at any given radius by the functional fit
\begin{equation}
\alpha(n) = \frac{a}{n} + b,
\label{simple_sersic}
\end{equation}
where for $r = (0.5,1,2,4)R_e$ the parameters are approximated by:
\begin{equation}
\begin{split}
a &=-1.54,-0.86,0.98,4.08,\\
b &=2.87,2.98,2.9,2.69.
\end{split}
\end{equation}
\smallskip
We provide $\alpha(n)$ separately for the bulge and for whole galaxies.
We measure the Sérsic indices $n$ for the bulge by fitting a two-component profile to the surface density profile of the stars, where the Sérsic index for the outer disc part is kept constant at $n_\mathrm{out}=1$.
Secondly we measure global Sérsic indices by fitting a single Sérsic component to the stellar surface-density profiles using \autoref{Sersic_formula}.

The values that we report in the following are measured face-on, i.e., the z-axis of the simulations are aligned with the AM-vector of the stars.
We fit the logarithm of the stellar surface-density profile of each simulated galaxy using a least-square minimisation.
We use logarithmically spaced bins from $100$pc to 0.1 $R_\mathrm{vir}$. Of all snapshots less then $7$ per cent did not converge for the two-component fit and less then $5$ per cent did not converge for the single-component fit. Obtained values for $R_e$ from the global fit are in agreement with the true half-mass radius with deviations of less then 9 per cent.

The measured values of $\alpha_\mathrm{v}$ and $\alpha_{\rho}$ as a function of Sérsic index $n$ are shown in \autoref{fig:all_n_combo}\hyperref[fig:all_n_combo]{a} for the bulge and \ref{fig:all_n_combo}\hyperref[fig:all_n_combo]{b} for the global fit.
There is good agreement between $\alpha_\mathrm{v}$ and $\alpha_\rho$ indicating Jeans equilibrium for all radii in the bulge and the disc as discussed above. Furthermore, measured values agree well with the theoretical expectations. Only for systems with small $n$ we find small discrepancies at radii $\geq 2 R_e$. This may be because at large radii there are deviations from spherically symmetric density distributions to more prolate systems.
With \autoref{simple_sersic} for $\alpha(n)$, the dynamical mass is recovered within 20\%.
It is apparent that for galaxies with $n\geq2$ the value for $\alpha$ is $\simeq 2.5-2.7$ at the effective radius.
\subsection{Prescriptions and Tests}
We approximate the values of $\alpha$ separately for galactic discs and for whole galaxies as a function of either $M_\star$ or $R/R_e$ with parametric third order polynomials of the form
\begin{equation}
\alpha(x) = \mathrm{a} x^2 + \mathrm{b}x + \mathrm{c}
\end{equation}
where $x=\log(M_\star/10^{9.5} \mathrm{M_\odot})$ or $x=R/R_e-1$ and $R_e$ is the half-mass radius of the analysed component.
The coefficients for the different components are given in \autoref{tab:alpha}. We estimate the dynamical mass using these prescriptions together with \autoref{eq1} and compare it to the true value of the dynamical mass in the simulations.
From the distribution of the ratio between the estimated dynamical mass and the true dynamical mass $M_\mathrm{dyn,estimated}/M_\mathrm{dyn,true}$ we can estimate the statistical $1\sigma$ confidence interval within which our prescriptions will successfully recover the dynamical mass. For the stars the typical scatter is $\lesssim 16\%$ for all prescriptions. For the gas the scatter is larger but above the threshold mass and at $R_\mathrm{e,gas}$ the error is $\lesssim 20\%$.

\begin{figure}
    \centering
    \includegraphics[width=\columnwidth]{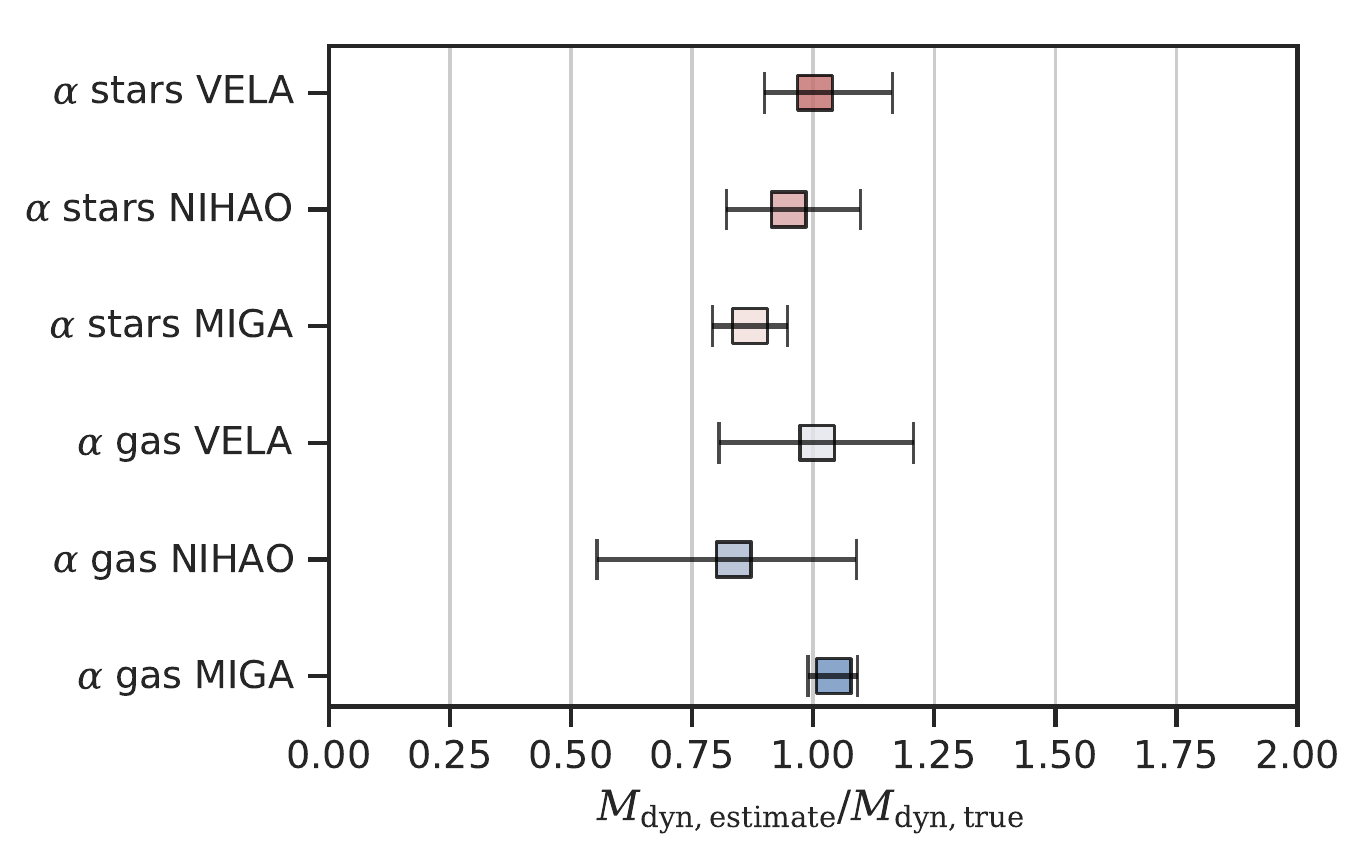}
    \caption{
    Accuracy of the dynamical mass prescription for decomposed kinematics in different simulation suites.
    Relative errors in terms of the ratio between the estimated dynamical mass $M_\mathrm{dyn,estimate}$ and the true dynamical mass $M_\mathrm{dyn,true}$ are shown using the prescriptions for $\alpha$ for whole galaxies with symbols for the median and error-bars for the $1\sigma$ confidence interval.
    The average dynamical mass is recovered to better than $20\%$ in all cases.
    }
    \label{fig:alpha_tests}
\end{figure}

Furthermore, we validate the prescriptions for whole galaxies by applying them to two different sets of zoom-in simulations, namely the MIGA suite detailed in Appendix \ref{app_ramses} and the NIHAO suite detailed in Appendix \ref{app_nihao}.
These simulations use different codes, numerical resolutions and models for galaxy formation physics where in particular different implementations for star formation and stellar feedback are used.
The resulting relative errors are shown in \autoref{fig:alpha_tests}.
Using the stars, we find that the estimated dynamical mass is slightly underestimated by $\lesssim15\%$ in MIGA and NIHAO.
Using the gas in NIHAO, the estimated dynamical mass is underestimated by a systematic bias of $\sim 20\%$. This is likely because NIHAO galaxies have more radial outflows caused by strong supernova feedback.
Using the gas in the MIGA galaxies, the estimated dynamical mass is overestimated by $\sim 5\%$.
For all simulations sets, the true value for the dynamical mass is recovered within the 1$\sigma$ scatter except for the stars in MIGA where it is recovered within 2$\sigma$. 
The average dynamical mass is recovered to better than $20\%$ in all cases which demonstrates the validity of our prescriptions.

\begin{table}
\caption{Coefficients for polynomial approximations to the different measured values of $\alpha$ for gas and stars in different volumes. $\alpha(x) = \mathrm{a} x^2 + \mathrm{b}x + \mathrm{c}$, where $x$ is the observable (where $m\equiv\log(M_\star/10^{9.5}) \mathrm{M_\odot})$ together with corresponding $1\sigma$ confidence intervals, where the number in brackets is for $M_\star > 10^{9.5} \mathrm{M_\odot}$ (the radial profiles were analysed above the threshold mass).}
\label{tab:alpha}
\begin{tabular}{lllrrrr}
\hline
$x$ & type & volume & a & b & c & $\Tilde{\sigma}[\%]$\\
\hline
$m$ & stars & disc & -0.078 & 0.411 & 2.304 & 16 (11)\\
$R/R_e - 1$ & stars& disc & -0.075 & 0.591 & 2.544 & 16 \\
$m$ & gas &disc & -0.298 & 0.061 & 1.411 & 37 (18)\\
$R/R_e - 1$ & gas & disc & -0.146 & 1.204 & 1.475 & 40 \\
\hline
$m$ & stars & sphere & -0.047 & 0.356 & 2.412 & 12 (9)\\
$m$ & gas & sphere & -0.243 & -0.147 & 1.486 & 20 (15)\\
\hline
\end{tabular}
\end{table} 

\section{Measuring the dynamical mass from the los velocity dispersion}
\label{sec:K}
When only a line-of-sight velocity dispersion $\sigma_l$ is available, the dynamical mass can be estimated using the virial factor $K$.
\subsection{The Virial Factor}
\label{sec_k_factor}
\begin{figure*}
  \includegraphics[width=.8\textwidth]{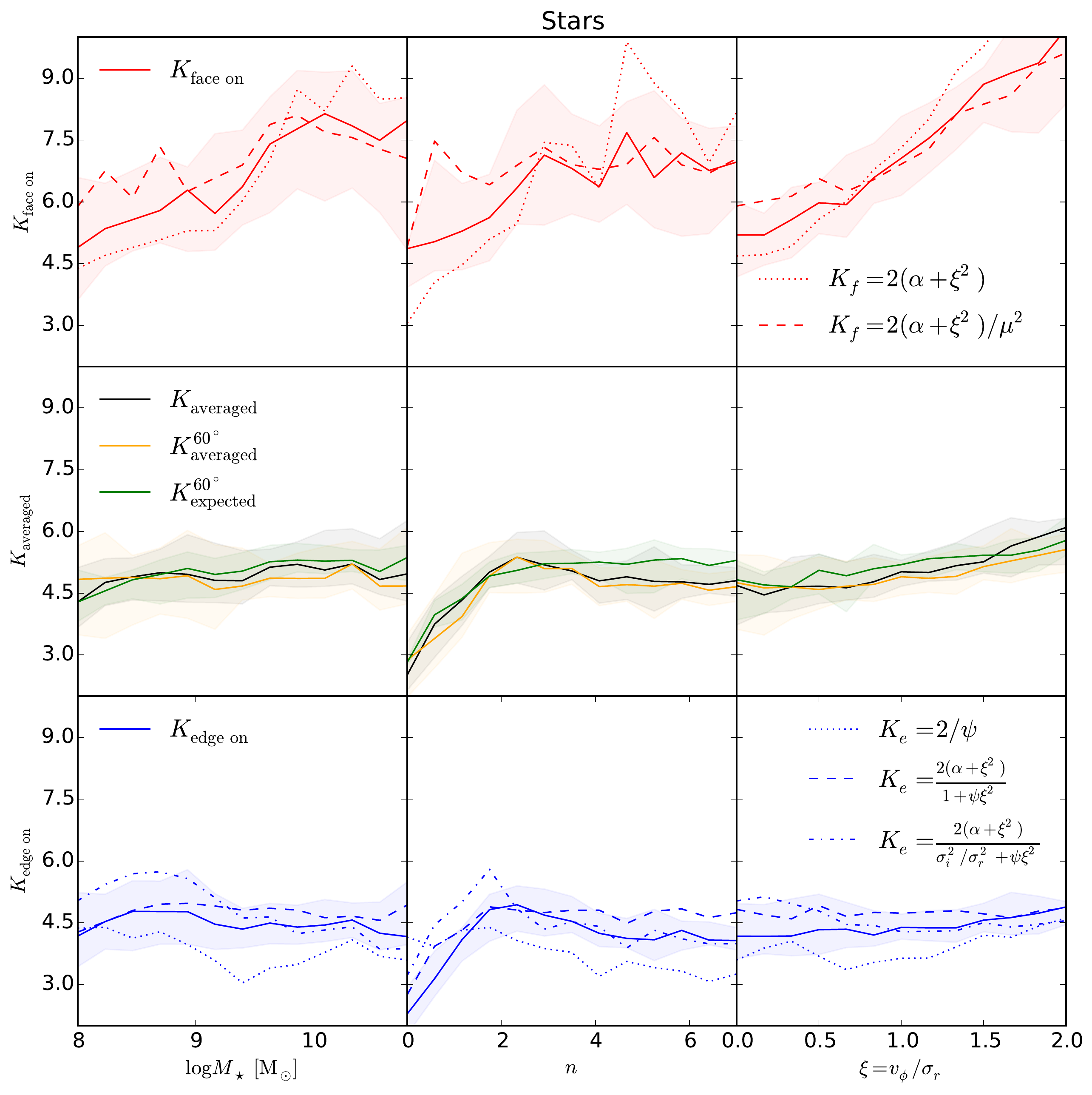}
  \caption{Measured values of the virial factor $K$ for the stars. The inclination changes from top to bottom: face-on (red), average (black) and edge-on (blue). Different columns show $K$ as a function of $M_\star,\, n$ and $v_\phi/\sigma_r$. Measured values are shown as solid lines with errors as shaded areas. Approximations are shown in dashed lines. Face-on, the model with anisotropy ($\mu$) successfully predicts $K$. For systems with low masses or small $v_\phi/\sigma_r$ the model without $\mu$ for $K_f$ is good.
  Edge-on, $K_e=2/\psi$ is not a good approximation since the systems are not purely rotation dominated. The other two models predict $K_e$ more successfully. The last model, which includes anisotropic dispersion, does not give a better approximation.
  Additionally in the middle row, $K$ is measured at a fixed angle of $60^\circ$ in yellow ($K^{60^\circ}_\mathrm{averaged}$) and compared to the theoretical expectations at this orientation in green ($K^{60^\circ}_\mathrm{expected}$). The similarity of the measured and predicted values for a fixed angle demonstrates the validity of our approach for $K$.
  In the face-on case $K$ increases with $M_\star$ and $n$ from 4.5 to 7.5, and with $\xi$ from 4.5 to 10. In the edge-on case $K$ is $\sim4.5$ except for $n<2$. Averaged over inclinations, $K$ is consistent with $K=5$ but only for $n>2$ and $0.25 < \xi < 1.5$.
}
  \label{fig:K_stars}
\end{figure*}
\begin{figure*}
  \includegraphics[width=.8\textwidth]{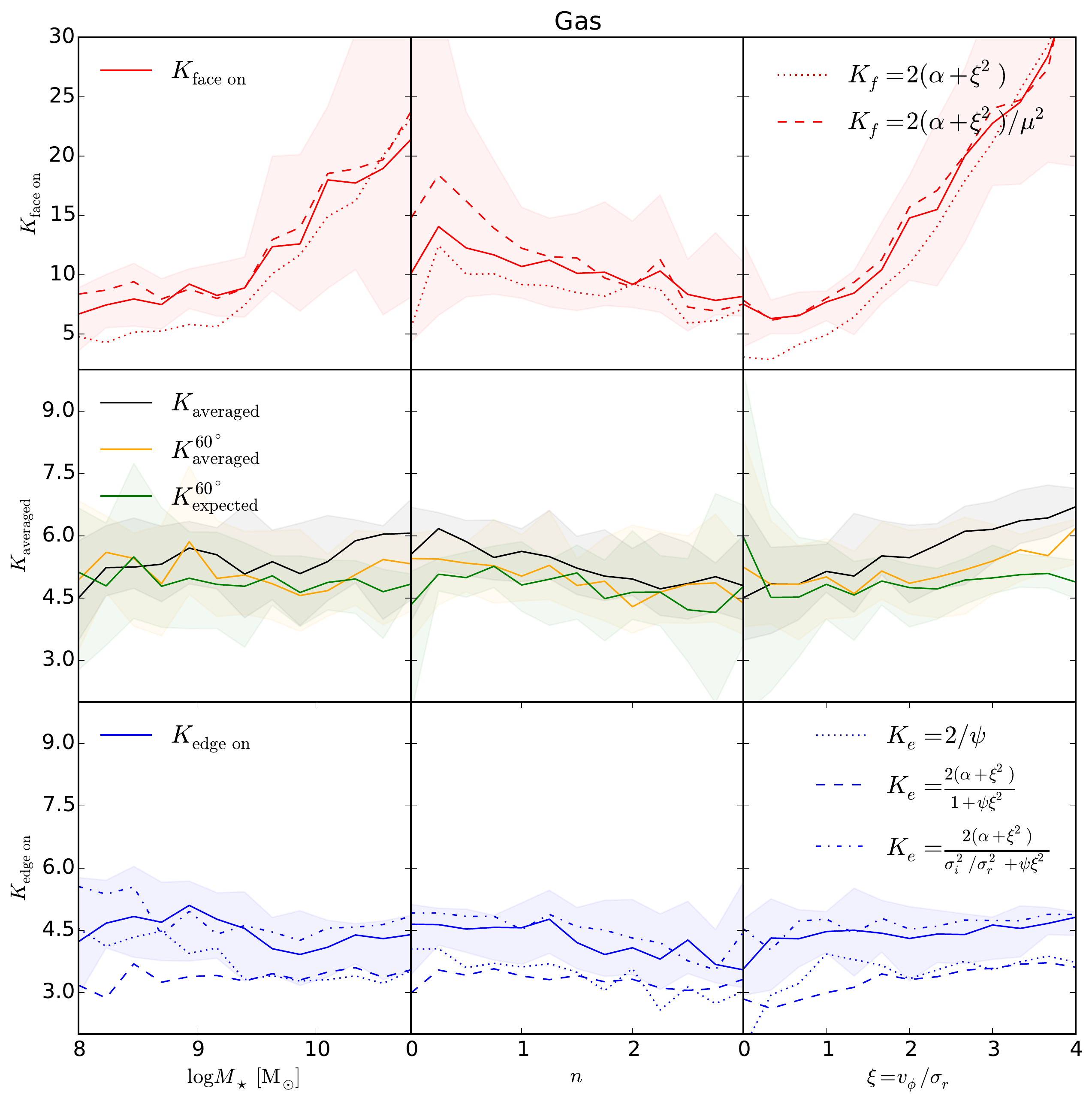}
  \caption{Measured values of the virial factor $K$ for the gas. The inclination changes from top to bottom: face-on (red), average (black) and edge-on (blue). Note the different scaling of the y-axis in the top row (face-on).
  Different columns show $K$ as a function of $M_\star,\, n$ and $v_\phi/\sigma_r$. Measured values are shown as solid lines with errors as shaded areas.
  Approximations are shown in dashed lines. In the face-on case, the two models give good approximations to the data, especially if accounted for anisotropic dispersion ($\mu$). Edge-on, only the last model, which accounts for anisotropic dispersion successfully predicts $K_f$. Additionally in the middle row, $K$ is measured at a fixed angle of $60^\circ$ in yellow ($K^{60^\circ}_\mathrm{averaged}$) and compared to the theoretical expectations at this orientation in green ($K^{60^\circ}_\mathrm{expected}$). The agreement between the measured and predicted values for a fixed angle demonstrates the validity of our approach for $K$.
  In the face-on cases $K$ rises with $M_\star$ and $\xi$ from $5$ to $30$ and decreases with $n$ from $12$ to $7.5$. Averaged over inclinations, $K$ increases with $M_\star$ and $\xi$ from $4.5$ to $6$, and decreases with $n$ from 6 to 4.5. Averaged over inclinations, $K$ is consistent with $K=5$ only for $n>1.5$ and $\xi <2$.
  Edge-on, the typical value for $K$ is $4.5$.}
\label{fig:K_gas}
\end{figure*}
\begin{figure*}
  \includegraphics[width=.8\textwidth]{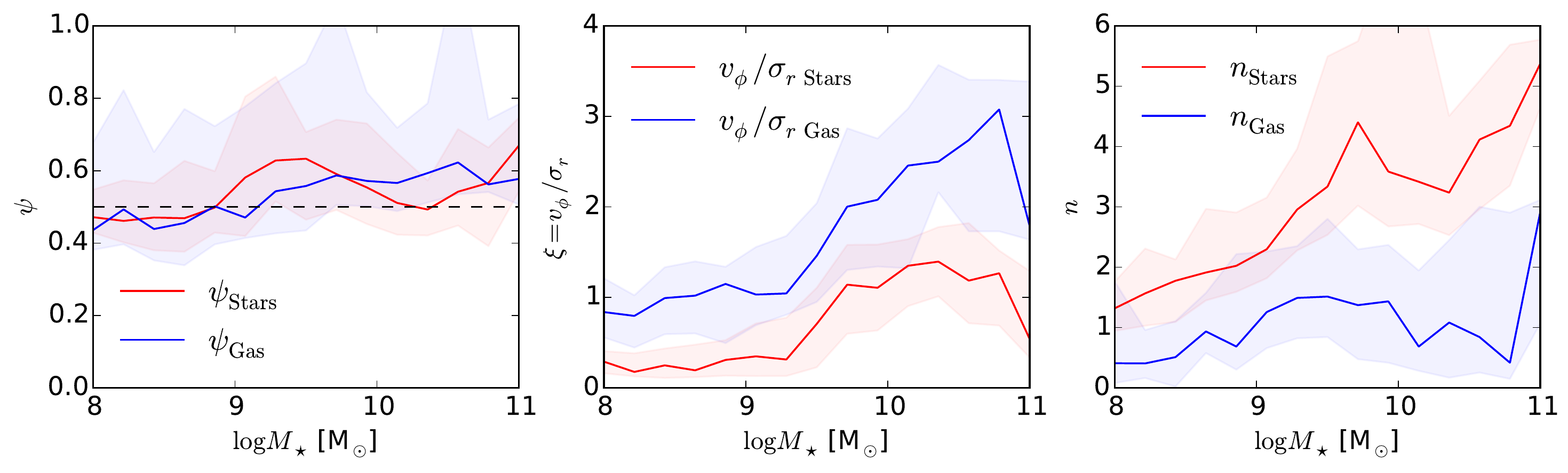}
  \caption{The line-of-sight projection $\psi$, the velocity ratio $\xi=v_\phi/\sigma_r$ and the Sérsic index $n$ for the stars and the gas separately evaluated at the effective radius $R_e$ of the analysed component as a function of the stellar mass. These values were used to calculate the virial factor $K$. The dashed line in the left plot represents $\psi = 0.5$ which is the value obtained from a flat rotation curve. Low-mass systems have flat gas and stellar rotation curves. High-mass systems $(M_\star > 10^{9.5}\mathrm{M_\odot})$, post-compaction, have slightly decreasing rotation curves. The stellar systems evolve from being highly dispersion dominated to systems with a dominant bulge and a disc component. The gas discs evolve from being marginally rotation dominated to highly rotation dominated with exponential profiles.}
  \label{fig:all_quan}
\end{figure*}
\subsubsection{Isotropic Case}
In order to estimate dynamical masses from the line-of-sight velocity dispersion $\sigma_l$ alone, the virial factor $K$ in \autoref{mdyn_k} should be evaluated.
We define the dynamical mass $M_\mathrm{dyn}$ to be twice the total mass inside the effective radius $R_e$, which using \autoref{gm/r} gives

\begin{equation}
M_\mathrm{dyn}=2 G^{-1} R_e V_c^2.
\end{equation}
We wish to express this in terms of observables, namely the effective radius $R_e$ and the line-of-sight (los) velocity dispersion $\sigma_l$,
\begin{equation}
M_\mathrm{dyn}=K G^{-1} R_e \sigma_l^2
\end{equation}
and find the value of the virial factor $K$ under different conditions, 
\begin{equation}
K = 2 \frac{V_c^2}{\sigma_l^2} = \frac{G M_\mathrm{dyn}}{R_e\sigma_l^2}.
\label{K_def}
\end{equation}

\smallskip
For a given line-of-sight with an inclination $i$ relative to the rotation axis ($i = 0$ for the face-on view) of the observed galaxy, the los velocity dispersion can be written as:
\begin{equation}
\sigma_l^2=\psi (\sin{i}\,v_\phi)^2 + \sigma_i^2,
\label{los_eq}
\end{equation}
where $\psi$ represents the projection of the rotation velocity at each point onto the direction of the line-of-sight. 
$\sigma_i$ is the actual velocity dispersion while $\sigma_l$ includes additional contribution from the rotation in the foreground and the background with respect to the galaxy centre line-of-sight distance to the line-of-sight dispersion.
For a transparent cylindrical disc extending to radius $R$ and viewed with an aperture of radius $R$,
\begin{equation}
\psi = \frac{1}{2}\int_0^R v_\phi^2(r)\Sigma(r) \diff r / \int_0^R \Sigma(r) \diff r,
\label{psi_def}
\end{equation}
where $v_\phi$ is expressed in units of the characteristic value $v_\phi$ that appears in \autoref{los_eq}. The value of $\psi$ depends on the shape of the rotation curve. For a flat rotation curve $\psi=1/2$ independent of $\Sigma(r)$. For an exponential disc with a rising rotation curve $v_\phi \propto r^{1/2}$ evaluated at the exponential radius $R_\mathrm{exp}$ gives $\psi\simeq 0.3$ and at $R\gg R_\mathrm{exp}$ it approaches unity. With a decreasing rotation curve $v_\phi \propto r^{-1/2}$, the value at $R_\mathrm{exp}$ is $\psi\simeq 1.2$ and at $R\gg R_\mathrm{exp}$ it approaches $\psi=1/2$.

In the face-on view $\sigma_i=\sigma_z$ and in the edge-on view $\sigma_i$ is an average of $\sigma_r$ and $\sigma_\phi$. In the simplest case we assume isotropy such that $\sigma_i$ is the same for every $i$.

Combining \autoref{eq:alpha}, \ref{K_def} and \ref{psi_def} and with
\begin{equation}
\xi \equiv \frac{v_\phi}{ \sigma_r},
\end{equation}
we obtain both for gas and stars separately the virial factor:
\begin{equation}
K = \frac{2 (\alpha + \xi^2)}{1 + \psi \xi^2 (\sin i )^2},
\label{eq:K_theo}
\end{equation}
with $\alpha$ from the pressure term in \autoref{eq:alpha}.
We can find approximations for various cases.
For example, in the dispersion-dominated case ($\xi = 0$) we have $K = 2 \alpha$,
which for a de Vaucouleurs spheroid ($n=4$) is $K=5.54$.
In the face-on case we obtain
\begin{equation}
K = 2 (\alpha + \xi^2),
\label{eq:face-on-app}
\end{equation}
for rotation-dominated systems ($\xi^2 \gg \alpha$) this becomes $K=2\xi^2$, which may yield values up to $\sim 30$.
In the edge-on case 
\begin{equation}
K = \frac{2 (\alpha + \xi^2)}{1 + \psi \xi^2},
\label{eq:edge-on-app}
\end{equation}
which for rotation dominated systems becomes $K=2/\psi$.

\subsubsection{Anisotropic Case}
In practice, the assumption of isotropic velocity dispersion may need to be modified, especially for the gas. 
This may represent a true anisotropy in the velocity dispersion, e.g., due to small-scale anisotropy in feedback-driven outflows or recycled inflows.
Alternatively, $\sigma_z$ as measured in the selected face-on view may include additional contributions from coherent motions.
Such motions can arise from not dealing with the best face-on view of the gas, which may occur if the face-on view is determined from the stars, and the gas and stellar discs are misaligned. A similar contamination can arise from intrinsic misalignment between the inner and outer disc, in the form of strong warps \citep{2015MNRAS.449.2087D}. In those two cases, gas rotation also contributes to $\sigma_z$. 

Without assuming isotropic dispersion \autoref{eq:K_theo} is replaced by
\begin{equation}
K = \frac{2 (\alpha + \xi^2)}{\sigma_i^2 / \sigma_r^2 + \psi \xi^2 (\sin i )^2},
\label{eq:K-anisotropic}
\end{equation}
where in the edge-on view $\sigma_i$ is an average of $\sigma_r$ and $\sigma_\phi$ and in the face-on view $\sigma_i=\sigma_z$.
We can thus generalise the toy model for the case
\begin{equation}
\mu \equiv \frac{\sigma_z}{\sigma_r},
\label{eq:nu}
\end{equation}
where for the face-on view the isotropic $K$ should be divided by $\mu^2$.

\subsection{Measuring \texorpdfstring{$K$}{K}}
Using \autoref{K_def} we can directly measure $K$ in the simulations.
We measure line-of-sight velocities along cylinders of radius $R_e$ and depth $3.4 R_e$, where the cylinder is rotated to $64$ randomly chosen orientations.
For each orientation, the line-of-sight velocities is calculated as 
\begin{equation}
\sigma_l = \sqrt{\langle v_z^2\rangle-\langle v_z\rangle^2},
\end{equation}
where $v_z$ is the velocity-component of the analysed component along the $z-$axis of the cylinder. Finally, the average over the $64$ orientations is computed.

\subsection{The Virial Factor in the Simulations}
We measure the virial factor $K$ separately for stars and gas using \autoref{K_def}. From the discussion in \autoref{sec_k_factor} it is apparent that the orientation of a galaxy influences the value for $K$ significantly. Therefore we separately measure $K$ for each snapshot in the face-on and edge-on cases as well as an average over $64$ random orientations.
Additionally, we measure $K$ at a fixed angle of $60^\circ$ and compare it to the expected values.
The average values for the measured $K$ as a function of the stellar mass of the system, the Sérsic index and $v_\phi/\sigma_r$, are shown in \autoref{fig:K_stars} for the stars and \autoref{fig:K_gas} for the (cold) gas.
We do not show an explicit redshift dependence of $K$ but instead note that $v_\phi/\sigma_r$ and $M_\star$ are strongly correlated with $z$ and can therefore be interpreted as indicators of the evolutionary phase.

The measurement is done at the effective radius $R_e$ of the analysed component, where the reference dynamical mass is taken to be twice the total mass inside a sphere of radius $R_e$. The line-of-sight velocity is measured along cylinders of radius $R_e$ and depth $3.4 R_e$. For the average case the cylinder is rotated to $64$ randomly chosen orientations.

To compare with theoretical values and approximations discussed in \autoref{sec_k_factor}, we use $\alpha, \psi$ and the velocity components evaluated at $R_e$ inside the same volume where $K$ is measured. The values for $\psi, \xi$ and $n$ are shown in \autoref{fig:all_quan}. The values for $\alpha$ are taken from \autoref{sec:alpha_all}.

\subsection{\texorpdfstring{$K$}{K} for the Stars}
Generally for the stars, we learn from \autoref{fig:all_quan} that most systems have Sérsic indices $n\sim1-2$, for $M_\star < 10^{9.5}\mathrm{M_\odot}$ and $n\sim2-6$ for $M_\star > 10^{9.5}\mathrm{M_\odot}$.
Pre-compaction, the ratio $v_\phi/\sigma_r < 0.5$ which increases post-compaction for more disc-like systems to $v_\phi/\sigma_r = 1.3 \pm 0.4$ (\autoref{fig:all_quan}).
This implies that for low-mass systems the stellar components of galaxies are not discs and that massive galaxies do have a disc component, but the bulge dominates.
It is apparent that $\psi\simeq 0.5$ for low-mass systems, implying a flat rotation curve for the stars. In galaxies above the threshold mass we find $\psi \sim 0.5 - 0.7$ which implies that these systems have slightly decreasing rotation curves.

From \autoref{fig:K_stars} the strong dependence of $K$ on the inclination is apparent. For high mass systems $K_\mathrm{stars}$ is $\sim 4.5$ in the edge-on case and $\sim 7.5$ in the face-on case. Furthermore, $K$ in the face-on case is increasing with mass and with $v_\phi/\sigma_r$. The reason for this is that high-mass systems form thin discs with large $v_\phi/\sigma_r$ such that the line-of-sight velocity in the face-on case gets smaller. Note that only in the average case $K\sim 5$, which is the value often used in the literature.
However, $K\sim 5$ is only consistent with our measurements for $n>2$ and for $0.25 < \xi < 1.5$, namely compact and thick discs.
In a similar study, \cite{2017MNRAS.469.2184F} demonstrated that for compact galaxies $K$ is systematically smaller than $5$ which is consistent with our findings.

The measured values in the face-on case agree well with the approximation $K_\mathrm{f}=2(\alpha + \xi^2)/\mu^2$ which accounts for anisotropic dispersion $\mu$. In the middle row, the value for the averaged $K$ is shown together with the measured value for galaxies oriented at $60^\circ$ in yellow ($K^{60^\circ}_\mathrm{averaged}$) and the expectation from \autoref{eq:K_theo} in green ($K^{60^\circ}_\mathrm{expected}$). The agreement between the average $K$ and the measured $K^{60^\circ}_\mathrm{averaged}$ is consistent with the fact that the average orientation is near $60^\circ$. Furthermore, the agreement between the averaged measured $K^{60^\circ}_\mathrm{averaged}$ and the expected $K^{60^\circ}_\mathrm{expected}$ demonstrates the validity of \autoref{eq:K_theo}.
In the edge-on case the approximations $K = 2 (\alpha + \xi^2)/(1 + \psi \xi^2)$
(\autoref{eq:edge-on-app}) and $K = 2 (\alpha + \xi^2)/(\sigma_i^2 / \sigma_r^2 + \psi \xi^2)$ (\autoref{eq:K-anisotropic}) yield good agreement with the measured values. Since the systems are not heavily rotation dominated, $K=2/\psi$ is not a good approximation.

\cite{2014RvMP...86...47C} compared predictions for the virial factor from three different formulas 
to various idealised models (see their table 2). These formulas should be used a priori for average inclinations and we therefore limit our comparison to this case.
We find that the formulas from \cite{1969ApJ...158L.139S} and \cite{2010MNRAS.406.1220W}\footnote{We have doubled the original values to convert from the half-light mass to the dynamical mass, ignoring variations in the mass-to-light ratio.}, where $K=7.5$ and $K=8.0$ respectively,
over-predict the dynamical mass.
The predicted value of $K=5$ from \cite{2006MNRAS.366.1126C} yields the closest approximation of the three formulas to our measured values, which is only valid for compact and thick discs as discussed above.
The values for $K$ from \cite{2010MNRAS.406.1220W} are larger than those from \cite{2006MNRAS.366.1126C} or ours because $\sigma_l$ is measured inside $R_\mathrm{vir}$ instead of $R_e$.
We compare our values to the model of \cite{2014RvMP...86...47C} which features a Sérsic profile for the stars together with a fixed $m=6$ Einasto profile for the DM. We use the values that were obtained following the definitions of \cite{2006MNRAS.366.1126C} which are the closest to our definitions.
We see the same trend of $K$ decreasing with $n$. Their values however over-predict the dynamical mass by $30-50\%$.
The discrepancies likely originate from the fact that the simple models used fixed DM and stellar profiles, isotropic velocities ($\beta=0$) and no rotation.

\subsection{\texorpdfstring{$K$}{K} for the Gas}
We extend the study for $K$ to the gas in galaxies.
From \autoref{fig:all_quan} it is apparent that for systems with $M_\star < 10^{9.5}\mathrm{M_\odot}$ the gas typically has $v_\phi/\sigma_r = 0.9 \pm 0.3$. For high-mass systems $(M_\star > 10^{9.5}\mathrm{M_\odot})$, post-compaction, thinner and more quiet gas discs are formed with $v_\phi/\sigma_r = 2.7 \pm 0.8$. The Sérsic index of the gas is typically $n\sim 0-1$, except during compaction when gas is driven to the centre into a compact object resulting in $n>1$.
It is apparent that $\psi\simeq 0.5$ for low-mass systems, implying a flat rotation curve for the gas. In galaxies above the threshold mass we find $\psi \sim 0.5 - 0.9$ which implies that these systems have decreasing rotation curves.

The dependence of $K$ on the inclination is even stronger for the gas compared to the stars. 
Note the different scaling of the y-axis in the first row of \autoref{fig:K_gas} compared to the other rows and compared to \autoref{fig:K_stars}.
For massive systems we find that $K_\mathrm{gas}$ is $\sim 4.5$ edge-on, $\sim 6$ for an average orientation and $\sim 10-30$ face-on.
We find that the literature value $K=5$ is only valid for the averaged case with $n>1.5$ and $v_\phi/\sigma_r < 2$, namely for compact and thick gas discs.

In the face-on case, $K$ is crudely approximated by $K_\mathrm{f}=2(\alpha + \xi^2)$ (\autoref{eq:face-on-app}). 
If anisotropic velocity dispersion are taken into account through $\mu=\sigma_z/\sigma_r$, the values in the face-on case are well approximated by $K_\mathrm{f}=2(\alpha + \xi^2)/\mu^2$ (\autoref{eq:K-anisotropic}).
This is expected since coherent motions, especially outflows from various feedback processes, can give rise to anisotropic gas velocity dispersions.

In the middle row of \autoref{fig:K_gas}, the measured value for galaxies orientated at $60^\circ$ in yellow ($K^{60^\circ}_\mathrm{averaged}$) and the expectation at $i=60^\circ$ in green ($K^{60^\circ}_\mathrm{expected}$) agree well, which demonstrates the validity of \autoref{eq:K_theo}.

In the edge-on case, the approximations for isotropic dispersion 
$K = 2 (\alpha + \xi^2)/(1 + \psi \xi^2)$ (\autoref{eq:edge-on-app})
underestimate $K$ by $\simeq 1.0$. However, there is good agreement if we account for anisotropic dispersion with $K = 2 (\alpha + \xi^2)/(\sigma_i^2 / \sigma_r^2 + \psi \xi^2)$ (\autoref{eq:K-anisotropic}).

\subsection{Prescriptions and Tests}
\begin{figure}
    \centering
    \includegraphics[width=\columnwidth]{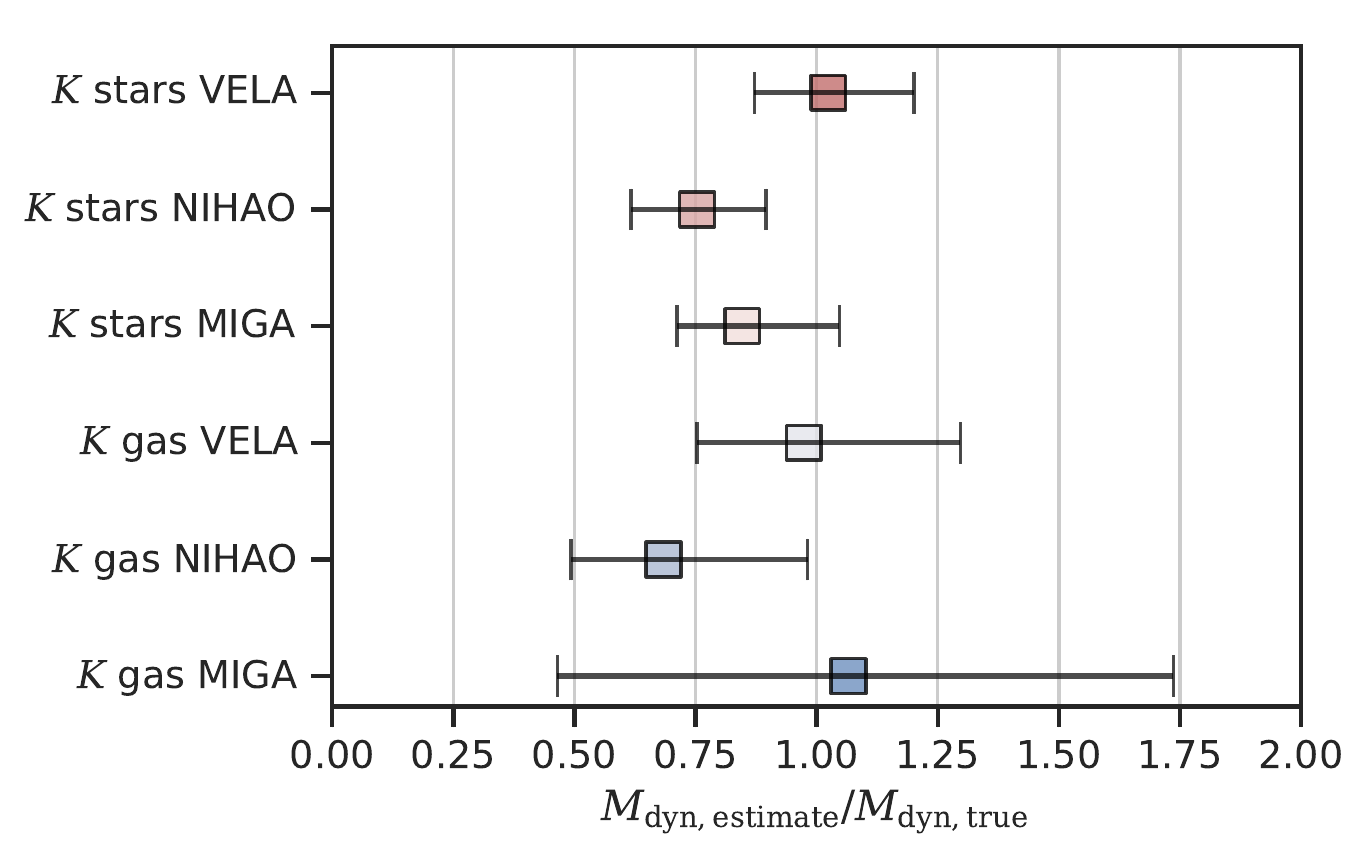}
    \caption{
    Relative errors on the dynamical mass using the prescriptions for $K$ for different simulations.
    The distributions for a given component and simulation set include values obtained using the prescriptions for all orientations and observables.
    Shown is the median and the 1$\sigma$ confidence interval.
    The dynamical mass is accurate to better than $20\%$ in MIGA.
    It is underestimated by $\sim35\%$ for the gas in NIHAO, possibly because of the strong feedback inducing outflows.
    }
    \label{fig:K_tests}
\end{figure}
We approximate the values of $K$ for the stars and the gas separately, as a function of either $M_\star$, $n$ or $\xi$ with parametric polynomials of the form
\begin{equation}
\alpha(x) = \mathrm{a} x^3 + \mathrm{b} x^2 + \mathrm{c} x + \mathrm{d}
\label{eq:param_form}
\end{equation}
where $x=\log(M_\star/10^{9.5} \mathrm{M_\odot})$, $n$, or $\xi$.
We fit $K$ in the face-on, average and edge-on case individually, using a least-square minimisation. 
For each case we fit separately a cubic, square, linear and constant polynomial form of \autoref{eq:param_form}. 
The best fit and the corresponding polynomial form is then chosen by evaluating the reduced $\chi^2$.
The coefficients for the different components are given in \autoref{tab:k}.

We estimate the dynamical mass using the prescriptions and compare it to the true value of the dynamical mass.
From the distribution of $M_\mathrm{dyn,estimated}/M_\mathrm{dyn,true}$ we can estimate the statistical $1\sigma$ confidence interval within which our prescriptions will successfully recover the dynamical mass.
Using these prescriptions the dynamical mass is recovered within relative errors of $\lesssim 20\%$ using \autoref{mdyn_k}. Only for the face-on cases using the gas, larger deviations occur. 

Furthermore, we test our prescriptions on two different sets of simulated galaxies (see Appendix \ref{app_ramses} for the MIGA suite and Appendix \ref{app_nihao} for the NIHAO suite).
\autoref{fig:K_tests} shows the resulting relative errors using the stars or the gas for different simulations with different feedback strengths.
Using $K$ for the stars, we find for all orientations and observables that the dynamical mass is underestimated by a systematic bias of $\sim25\%$.
The statistical error is comparable to the scatter obtained from the VELA simulations.
For the gas, using the prescriptions for $K$ in the NIHAO simulations under-predict the dynamical mass by $\sim 35\%$. However, using the prescriptions for $K$ in the MIGA simulations over-predicts the dynamical mass by $\sim 7\%$. The larger scatter for $K$ using the gas in NIHAO and MIGA originates mostly from face-on cases and is caused by stronger gas motions driven by the stronger feedback that is implemented in these simulations. The good agreement between the estimated and true dynamical masses in the different simulations demonstrates the accuracy of our prescriptions.

\begin{table}
\caption{Coefficients for polynomial approximations to the different measured values of $K$ for the gas and stars.
$K(x, \mathrm{a...d}) = \mathrm{a}x^3 + \mathrm{b}x^2 + \mathrm{c}x + \mathrm{d}$, 
where $x$ is the observable $(m = \log( M_\star / 10^{9.5} \mathrm{M_\odot}), n, \xi=v_\phi / \sigma_r)$ and relative $1\sigma$ confidence interval on the estimated value for $M_\mathrm{dyn}$.}
\label{tab:k}
\begin{tabular}{llrrrrrrr}
\hline
$K$ & $x$ & a & b & c & d & $\Tilde{\sigma}[\%]$\\
\hline
Stars\\
\hline
$K_\mathrm{f} $ & $m$ & 0.00 & 0.00 & 1.18 & 6.70 & 20\\
$K_\mathrm{f} $ & $n-4$ & 0.00 & -0.11 & 0.13 & 6.96 & 25\\
$K_\mathrm{f} $ & $\xi - 1$ & 0.00 & 0.00 & 2.53 & 7.23 & 16\\
$K_\mathrm{avg} $ & $m$ & 0.00 & 0.00 & 0.10 & 4.97 & 13\\
$K_\mathrm{avg} $ & $n -4$ & 0.05 & -0.07 & -0.31 & 5.09 & 13\\
$K_\mathrm{avg} $ & $\xi -1$ & 0.00 & 0.00 & 0.76 & 5.04 & 15\\
$K_\mathrm{e} $ & $m$ & 0.00 & 0.00 & -0.14 & 4.45 & 15\\
$K_\mathrm{e} $ & $n -4$ & 0.05 & -0.04 & -0.36 & 4.46 & 13\\
$K_\mathrm{e} $ & $\xi -1$ & 0.00 & 0.00 & 0.40 & 4.39 & 16\\
\hline
Gas\\
\hline
$K_\mathrm{f} $ & $m$ & 0.00 & 2.80 & 5.73 & 10.17 & 65\\
$K_\mathrm{f} $ & $n - 1$ & 0.00 & 0.73 & -3.55 & 11.67 & 76\\
$K_\mathrm{f} $ & $\xi - 1$ & 0.00 & 2.43 & 2.50 & 7.68 & 46\\
$K_\mathrm{avg} $ & $m$ & 0.00 & 0.00 & 0.26 & 5.49 & 20\\
$K_\mathrm{avg} $ & $n - 1$ & 0.00 & 0.00 & -0.52 & 5.61 & 18\\
$K_\mathrm{avg} $ & $\xi - 1$ & 0.00 & 0.00 & 0.44 & 5.26 & 19\\
$K_\mathrm{e} $ & $m$ & 0.00 & 0.00 & -0.24 & 4.52 & 19\\
$K_\mathrm{e} $ & $n - 1$ & 0.00 & 0.00 & -0.32 & 4.37 & 19\\
$K_\mathrm{e} $ & $\xi - 1$ & 0.00 & 0.00 & 0.18 & 4.18 & 20\\
\hline
\end{tabular}
\end{table}

\section{Conclusions}
\label{sec:conc}
We derived recipes for evaluating the dynamical mass of a galaxy from kinematic measurements, based on the VELA suite of high-resolution zoom-in cosmological simulations.
First, we studied the validity of Jeans and hydrostatic equilibrium for the stars and gas respectively.
Assuming cylindrical symmetry and a constant, isotropic velocity dispersion $\sigma_r$, the Jeans equation was written as
\begin{equation}
V_c^2 = v_\phi^2 + \alpha\, \sigma_r^2
\label{jeans_in_sum}
\end{equation}
with the circular velocity $V_c$, the rotational velocity $v_\phi$ and $\alpha=-\diff \ln \rho / \diff \ln r$ the logarithmic slope of the density profile.
We have compared values for $\alpha_\rho$ obtained from the density slope (\autoref{alpha_def}) and $\alpha_\mathrm{v}$ obtained from the velocities (\hyperref[alpha_v_def]{Equations} \ref{alpha_v_def} and \ref{jeans_in_sum}). Equilibrium is indicated if both measurements yield consistent values. Otherwise we inspect corrections from aspherical potentials, non-constant and anisotropic velocity dispersions.
Our analysis reveals that equilibrium is valid for stars and gas above a stellar mass of $M_\star \sim 10^{9.5} \mathrm{M}_\odot$,
the threshold mass for long-lived discs associated with infrequent merger-driven spin flips \citep{2020MNRAS.493.4126D}, 
compaction events \citep{2015MNRAS.450.2327Z,2016MNRAS.457.2790T,2016MNRAS.458..242T,2016MNRAS.458.4477T,2020MNRAS.496.5372D}
and less effective supernova feedback \citep{1986ApJ...303...39D,2015MNRAS.453..408C,2019MNRAS.488.4753D}.
The equilibrium is typically valid out to $\sim 5 R_\mathrm{e}$.
We separately analysed the bulge component, the disc component and the two components together.
For each component we provide functional fits for $\alpha$ as a function of mass and radius to enable measurements of the dynamical mass $M_\mathrm{dyn}$ from stellar and gas kinematics.
When only the line of sight velocity dispersion $\sigma_l$ within $R_\mathrm{e}$ is available, we provide fitting functions for the virial factor $K$ which was defined as
\begin{equation}
    K = \frac{G M_\mathrm{dyn}}{R_e\sigma_l^2},
    \label{K_again}
\end{equation}
as a function of various observables.
A summary of our results is as follows: 

\begin{itemize}
    \item For high-mass systems with $M_\star > 10^{9.5} \mathrm{M}_\odot$, the stars are in fair Jeans equilibrium.
    We find the typical value to be $\alpha\simeq2.6$ for the stars at the effective radius.
    The effect of a non-constant velocity dispersion ($\gamma \sim 0.3$) is counteracted by the effect of a non-spherical potential ($\Delta_Q \sim -0.3$), while the anisotropy has a negligible effect ($\beta \simeq 0$).
    
    \item For the gas we find that hydrostatic equilibrium is valid above a similar mass threshold with a typical value of $\alpha\simeq 1$. 
    We find that the corrections from anisotropic velocity dispersions ($\beta \simeq 0.5$) as well as contributions from non-spherical potentials and non-constant velocity dispersion ($\gamma \sim -0.2$, $\Delta_Q \sim -0.5$) are not negligible.
    
    \item For galaxies below the mass threshold, we find small but systematic deviations from Jeans and hydrostatic equilibrium. However, these deviations are sufficiently small to allow crude estimates of the dynamical mass.
    
    \item For massive galaxies, the equilibrium is valid up to radii of $\sim 5 R_\mathrm{e}$ where $\alpha$ increases from $\simeq 2.5$ below $R_\mathrm{e,stars}$ to $\simeq 3.5$ at $5R_\mathrm{e,stars}$ for the stars, and from $\simeq1$ below $R_\mathrm{e,gas}$ to $\simeq4$ at $5R_\mathrm{e,gas}$ for the gas.
    
    \item We find large deviations from the predictions for self-gravitating exponential discs. By analysing the force exerted by each component in the vertical direction at a distance $z=-H_d$ from the disc plane, we learned that the contribution from the spheroidal components (dark matter and bulge stars) dominate over the contribution from the disc component (gas and disc stars) by a factor larger than two at $R<R_\mathrm{e,gas}$, and by an even larger factor at large radii.
    
    \item We provided $\alpha$ as a function of Sérsic index. A simple estimator for the theoretical predictions is given by $\alpha(n) = \mathrm{a} / n + \mathrm{b}$, where we provide $\mathrm{a,b}$ at various radii.
    For an exponential disc, $\alpha(n=1)=2$ at the effective radius $R_\mathrm{e}$.
    For a de Vaucoleurs profile $\alpha(n=4) \simeq 2.7$ at $R_\mathrm{e}$.
    
    \item When only an estimate of the line-of-sight velocity dispersion $\sigma_l$ within $R_\mathrm{e}$ is available, we use the demonstrated validity of hydrostatic and Jeans equilibrium and provide the virial factor $K$ (\autoref{K_def} and \ref{K_again})
    for different inclinations and as a function of either $M_\star,n$ or $v_\phi/\sigma_r$. 
    For the stars, $K$ varies from $4.5$ to $7.5$ from edge-on to face-on views respectively. For the gas it varies from $4.5$ to $30$.
    
    \item For the stars, the standard value of $K=5$ is valid in the simulations for values averaged over inclinations, but only for $n>2$ and $0.25 < v_\phi/\sigma_r < 1.5$.
    Similarly, for the gas we find that $K=5$ is only valid for values averaged over inclinations with $n>1.5$ and $v_\phi/\sigma_r < 2$, namely for compact and thick gas discs.

\end{itemize}

\section*{Acknowledgements}
We acknowledge stimulating discussions with Andi Burkert, Reinhard Genzel and Romain Teyssier. This work was partly supported by the Minerva foundation, the Swiss National Supercomputing Center (CSCS) project s890 - ``Predictive models for galaxy formation'', the Swiss National Science Foundation (SNSF) project 172535 - ``Multi-scale multi-physics models of galaxy formation'' and the grants France-Israel PICS, I-CORE Program of the PBC/ISF 1829/12, BSF 2014-273 and NSF AST-1405962, GIF I-1341-303.7/2016, DIP STE1869/2-1, ISF 861/20, GE625/17-1. DC is a Ramon-Cajal Researcher and is supported by the Ministerio de Ciencia, Innovaci\'{o}n y Universidades (MICIU/FEDER) under research grant PGC2018-094975-C21.

\section*{Data Availability}
The data underlying this article will be shared on reasonable request to the corresponding author.

\bibliographystyle{mnras}
\bibliography{bib_adsabs}
\appendix
\section{The VELA Cosmological Simulations}
\label{sec:app_vela}

\begin{table*}
\centering
\scalebox{1.0}{
\begin{tabular}{@{}cccccccccccccccc}
\multicolumn{16}{c}{{\bf Properties of the VELA galaxies}} \\
\hline
Galaxy & $\Mv$ & $\Ms$ & $\Mg$ & SFR & $R_\mathrm{e,stars}$ & $\Rd$ & $\Hd$ & $v_\phi$ & $\sigma_r$ & $e$ & $f$ & $a_{\rm fin}$ & $n_\mathrm{bulge}$ & $r_\mathrm{e,in}$ & $B/T$\\
 & $10^{12}\msun$ & $10^{10}\msun$ & $10^{10}\msun$ & $\msun/\yr$ & kpc & kpc & kpc & km/s & km/s & & & & & kpc & \\

\hline
\hline
01 & 0.16 & 0.20 & 0.14 & 2.64 & 0.93 & 5.15 & 2.57 & 66.0 & 50.3 & 0.72 & 0.97 & 0.50 & 2.26 & 0.37 & 0.44 \\
02 & 0.13 & 0.16 & 0.12 & 1.43 & 1.81 & 6.37 & 3.57 & 71.6 & 39.1 & 0.81 & 0.98 & 0.50 & 2.58 & 0.81 & 0.50 \\
03 & 0.14 & 0.38 & 0.08 & 3.67 & 1.41 & 5.21 & 2.34 & 78.3 & 59.7 & 0.75 & 0.96 & 0.50 & 2.87 & 2.39 & 0.49 \\
04 & 0.12 & 0.08 & 0.08 & 0.45 & 1.73 & 5.71 & 2.79 & 23.3 & 54.3 & 0.96 & 0.88 & 0.50 & 2.38 & 0.56 & 0.49 \\
05 & 0.07 & 0.07 & 0.05 & 0.38 & 1.81 & 5.36 & 1.98 & 60.4 & 32.8 & 0.94 & 0.75 & 0.50 & 1.18 & 0.66 & 0.51 \\
06 & 0.55 & 2.14 & 0.33 & 20.60 & 1.05 & 2.53 & 0.42 & 221.1 & 48.1 & 0.56 & 1.00 & 0.37 & 4.44 & 1.03 & 0.50 \\
07 & 0.90 & 5.75 & 0.79 & 18.13 & 2.85 & 12.59 & 2.06 & 285.5 & 71.4 & 0.85 & 1.00 & 0.54 & 8.15 & 0.95 & 0.38 \\
08 & 0.28 & 0.35 & 0.15 & 5.70 & 0.74 & 4.03 & 1.53 & 91.6 & 48.5 & 0.80 & 0.96 & 0.57 & 6.69 & 0.39 & 0.50 \\
09 & 0.27 & 1.03 & 0.29 & 3.57 & 1.74 & 7.34 & 2.12 & 152.9 & 36.0 & 0.99 & 0.85 & 0.40 & 4.37 & 0.28 & 0.55 \\
10 & 0.13 & 0.60 & 0.13 & 3.20 & 0.46 & 4.51 & 1.19 & 137.1 & 40.9 & 0.50 & 0.99 & 0.56 & 7.50 & 0.34 & 0.48 \\
11 & 0.27 & 0.76 & 0.33 & 8.94 & 2.14 & 8.34 & 5.08 & 121.3 & 71.1 & 0.90 & 0.80 & 0.46 & - & 2.90 & 0.44 \\
12 & 0.27 & 1.95 & 0.20 & 2.70 & 1.13 & 6.53 & 1.72 & 181.2 & 43.7 & 0.97 & 0.78 & 0.44 & 3.81 & 0.21 & 0.50 \\
13 & 0.31 & 0.57 & 0.35 & 4.48 & 2.48 & 9.74 & 4.75 & 131.7 & 41.3 & 0.97 & 0.88 & 0.40 & 3.12 & 0.42 & 0.41 \\
14 & 0.36 & 1.26 & 0.44 & 23.31 & 0.32 & 1.10 & 0.14 & 213.9 & 82.8 & 0.43 & 0.98 & 0.41 & 5.12 & 0.22 & 0.46 \\
15 & 0.12 & 0.51 & 0.08 & 1.35 & 1.07 & 6.26 & 1.08 & 110.2 & 37.9 & 0.80 & 0.98 & 0.56 & 3.80 & 0.15 & 0.47 \\
16 & 0.50 & 4.09 & 0.50 & 18.46 & 0.61 & 6.05 & 0.99 & 269.4 & 104.8 & 0.37 & 0.98 & 0.24 & 2.00 & 0.25 & 0.44 \\
17 & 1.13 & 8.48 & 1.11 & 61.37 & 1.36 & 7.70 & 1.10 & 288.6 & 180.6 & 0.43 & 0.99 & 0.31 & 3.35 & 0.44 & 0.48 \\
19 & 0.88 & 4.49 & 0.57 & 40.46 & 1.22 & 1.55 & 0.12 & 257.2 & 91.9 & 0.70 & 0.99 & 0.29 & 1.38 & 0.35 & 0.34 \\
20 & 0.53 & 3.59 & 0.35 & 5.55 & 1.72 & 9.57 & 2.75 & 235.0 & 62.0 & 0.78 & 1.00 & 0.44 & 1.76 & 0.56 & 0.38 \\
21 & 0.62 & 4.05 & 0.43 & 7.89 & 1.73 & 9.48 & 1.18 & 261.6 & 42.9 & 0.52 & 1.00 & 0.50 & 2.60 & 1.31 & 0.29 \\
22 & 0.49 & 4.40 & 0.25 & 12.00 & 1.31 & 4.70 & 0.40 & 285.6 & 50.3 & 0.48 & 1.00 & 0.50 & 4.42 & 1.34 & 0.50 \\
23 & 0.15 & 0.76 & 0.13 & 3.06 & 1.16 & 6.28 & 1.54 & 133.0 & 49.2 & 0.78 & 0.99 & 0.50 & 4.43 & 0.80 & 0.50 \\
24 & 0.28 & 0.88 & 0.25 & 3.88 & 1.68 & 7.29 & 1.95 & 131.5 & 42.0 & 0.99 & 0.97 & 0.48 & 4.58 & 1.48 & 0.50 \\
25 & 0.22 & 0.69 & 0.08 & 2.29 & 0.73 & 5.70 & 0.82 & 93.9 & 71.3 & 0.80 & 0.99 & 0.50 & 3.19 & 0.99 & 0.50 \\
26 & 0.36 & 1.58 & 0.26 & 9.36 & 0.74 & 5.42 & 1.30 & 179.6 & 65.0 & 0.74 & 1.00 & 0.50 & 3.84 & 1.35 & 0.36 \\
27 & 0.33 & 0.71 & 0.29 & 6.10 & 1.98 & 9.16 & 4.97 & 122.8 & 60.4 & 0.25 & 0.98 & 0.50 & 3.91 & 0.85 & 0.35 \\
28 & 0.20 & 0.18 & 0.21 & 5.54 & 2.32 & 5.66 & 2.97 & 37.3 & 84.6 & 0.92 & 0.63 & 0.50 & 5.39 & 0.99 & 0.48 \\
29 & 0.53 & 2.00 & 0.49 & 11.83 & 1.46 & 8.50 & 1.41 & 195.2 & 77.7 & 0.93 & 0.27 & 0.50 & 4.47 & 2.50 & 0.45 \\
30 & 0.31 & 1.57 & 0.24 & 2.97 & 1.43 & 9.32 & 1.67 & 192.1 & 37.3 & 0.68 & 1.00 & 0.34 & 3.56 & 0.76 & 0.41 \\
31 & 0.23 & 0.78 & 0.13 & 15.26 & 0.43 & 4.19 & 0.96 & 195.4 & 48.1 & 0.82 & 0.99 & 0.19 & 3.19 & 0.28 & 0.52 \\
32 & 0.59 & 2.66 & 0.43 & 14.86 & 2.58 & 4.98 & 1.06 & 195.4 & 56.4 & 0.84 & 1.00 & 0.33 & 1.31 & 0.19 & 0.47 \\
33 & 0.83 & 4.81 & 0.44 & 32.68 & 1.23 & 4.59 & 0.88 & 262.7 & 114.2 & 0.49 & 0.95 & 0.39 & 3.14 & 0.49 & 0.50 \\
34 & 0.52 & 1.57 & 0.44 & 14.47 & 1.84 & 5.29 & 1.87 & 156.9 & 70.8 & 0.29 & 1.00 & 0.35 & 7.88 & 1.06 & 0.50 \\
35 & 0.23 & 0.56 & 0.25 & 22.93 & 0.33 & 1.13 & 0.30 & 204.4 & 40.4 & - & - & 0.22 & 2.48 & 0.18 & 0.58 \\
\hline
\end{tabular}
}
\caption{Relevant global properties of the VELA galaxies.
The quantities are quoted at $z=2$ ($a=0.33$) or at the final time-step $a_{\rm fin}$ when it is $<0.33$.
$\Mv$ is the total virial mass.
The following four quantities are measured within $0.1\Rv$:
$\Ms$ is the stellar mass, $\Mg$ is the gas mass, SFR is the star formation rate, and $R_\mathrm{e,stars}$ is the half-stellar-mass radius.
The disc outer cylindrical volume, as defined in \citet{2014MNRAS.443.3675M}, is given by $\Rd$ and $\Hd$, the disc radius and half height that contain 85\% of the gas mass within $0.15\,\Rv$.
$v_\phi$ and $\sigma_r$ are the rotation velocity and the radial velocity dispersion of the gas.
$e$ and $f$ are the shape parameters of the gas distribution, representing the ``elongation'' and ``flattening'' as defined in \citet{2016MNRAS.458.4477T}.
$a_{\rm fin}$ is the expansion factor at the last output. For the final snapshot the following three quantities are listed: $n_\mathrm{bulge}$ is the Sérsic index of the bulge, $r_\mathrm{e,in}$ is the effective radius of the bulge and $B/T$ is the bulge-to-total mass ratio.
}
\label{tab:sample}
\end{table*}
We base our analysis on cosmological zoom-in hydro-cosmological simulations.
We analyse the VELA simulation suite \citep{2014MNRAS.442.1545C, 2015MNRAS.450.2327Z} which consists of 34 simulated galaxies produced with the Adaptive Refinement Tree (\texttt{ART}) code \citep{1997ApJS..111...73K,2009ApJ...695..292C}.
In this appendix we give an overview of the key aspects of the simulations and their limitations.
Most of the galaxies reach a redshift of $z=1$. The analysis is done in steps of scale factor $\Delta a = 0.01$ which corresponds to $\Delta t \simeq 150 \, \mathrm{Myr}$ at redshift $z=2$.

The \texttt{ART} code uses high-resolution algorithms to calculate gravitational forces for N-body systems and solves the equations of hydrodynamics on a grid. The grid has a regular Cartesian structure and is split into smaller cells in dense regions. 

First, the initial conditions for the high-resolution hydrodynamical run are generated in low resolution N-body dark-matter-only simulations. Halos are selected based on their virial masses. Halos with major mergers at $z=1$ were excluded which may affect the resulting galaxies to be more disc-dominated.
The range of the selected halos for the hydrodynamical run was selected to be $M_\mathrm{vir} = (2 \times 10^{11} - 2 \times 10^{12} )\mathrm{M}_\odot$ \citep{2015MNRAS.450.2327Z, 2014MNRAS.442.1545C} where the virial mass is the total mass in a sphere of radius $R_\mathrm{vir}$ chosen such that it encompasses an overdensity of $\Delta (a) \simeq (18\pi^2-82 \Omega_\Lambda(a)-39\Omega_\Lambda(a)^2)/\Omega_\mathrm{m}(a)$ \citep{1998ApJ...495...80B, 2006MNRAS.368....2D} which is an approximation for a flat cosmological model with $\Omega_\Lambda(a) = 1- \Omega_\mathrm{m}(a)$.

After gas was inserted in the initial conditions, the galaxies have been evolved in an hydrodynamical run using additional subgrid physics on an adaptive co-moving mesh. The best physical resolution is $17-35\pc$ at all times, which is achieved at densities of $\sim10^{-4}-10^3\cmc$. The force resolution is two cells. Dark matter particles have a mass of $8.3 \times 10^4 \mathrm{M}_\odot$ and stars have a mass of $10^3 \mathrm{M}_\odot$. Refinement to a higher level of resolution is done if a cell contains a mass of dark matter and stellar particles larger than $2.6 \times 10^5 \mathrm{M}_\odot$ or if the gas mass is higher than $1.5 \times 10^6 \mathrm{M}_\odot$.

Additionally, the code contains a set of subgrid physics models that describe many relevant processes of galaxy formation that are not directly calculable because of the limited resolution \citep{2009ApJ...695..292C,2010MNRAS.404.2151C,2012MNRAS.420.3490C,2014MNRAS.442.1545C,2015MNRAS.450.2327Z,2014MNRAS.443.3675M}. Those processes include gas and metal cooling, photoionization heating, stochastic star formation, stellar feedback, metal enrichment, stellar mass loss, thermal feedback from supernovae, stellar winds, gas recycling and an implementation of feedback from radiation pressure as described in \cite{2014MNRAS.442.1545C}.

\section{The MIGA Cosmological Simulations}
\label{app_ramses}
The MIGA suite consists of 9 simulated galaxies that were performed with the adaptive mesh refinement (AMR) code \texttt{RAMSES} \citep{2002A&A...385..337T}. The simulation methods are described in detail in \cite{2020MNRAS.492.1385K} and \cite{2020MNRAS.497.4346K}.
We analyse 676 snapshots in the redshift range $z=0-5$.

From an $N$-body simulation with $512^3$ dark matter particles in a periodic box of size 25~$h^{-1}$Mpc, halos at $z=0$ with virial masses in the range $M_\mathrm{vir}=(0.75 - 1.5) \times 10^{12}\mathrm{M_\odot}$ were selected, where the virial radius was calculated using a spherical over-density according to the definition of \cite{1998ApJ...495...80B}.
Additionally we required that the halos were in relative isolation at $z=0$ and without major mergers events after $z=1$. We then performed zoom-in hydro-cosmological simulations where refinement levels were progressively released to enforce a quasi-constant physical resolution, such that the smallest cells have sizes $\Delta x_{\rm min}=55$pc.
The mass of dark matter particles is $m_{\rm dm}=2.0 \times 10^{5}$M$_\odot$ and the initial baryonic mass is $m_{\rm bar}=2.9 \times 10^{4}$M$_\odot$.

Star-formation is modelled using a Schmidt-law.
We use a novel approach for the star-formation efficiency per free-fall time $\epsilon_\mathrm{ff}$ where $\epsilon_\mathrm{ff}$ is based on the turbulent state of the gas. This allows for varying efficiencies ranging from $0\%$ to $100\%$, different to traditional models where $\epsilon_\mathrm{ff}=1\%$ is set to a constant value. Furthermore, in addition to thermal energy also momentum from supernova explosions is injected into the surrounding gas if the cooling radius is unresolved by the grid.
Individual supernovae explosions are resolved in time, with each star particle triggering multiple supernovae explosion, spread randomly between $3$ and $20$ Myrs after the birth of the star particle. Additionally, \texttt{RAMSES} contains a set of physics models the describe gas cooling and heating, metal cooling, heating by the UV background, stellar winds, photoionization, stellar mass loss and metal enrichment.

\section{The NIHAO Cosmological Simulations}
\label{app_nihao}
The NIHAO suite \citep{2015MNRAS.454...83W} consists of $\sim90$ cosmological zoom-in hydrodynamical simulations that were ran with the Smoothed Particle Hydrodynamics (SPH) code \texttt{gasoline2} \citep{2017MNRAS.471.2357W}.

The NIHAO sample consists of halo masses in the range of $\log{M_\mathrm{vir}/M_\odot} = 9.5 - 12.3$ chosen to be in isolation at $z=0$. In the selection of the halos, merging histories, concentrations and spin parameters were not taken into account.
The particle mass and force softening was chosen such that the mass profiles at 1 per cent of the virial radius are well resolved.
The code contains a set of subgrid physics models that describe the processes of turbulent mixing of metals and thermal energy \citep{2008MNRAS.387..427W}, metal cooling, heating by the UV background \citep{2010MNRAS.407.1581S} and ionising feedback from massive stars \citep{2013MNRAS.428..129S}.

Stellar feedback is modelled using the blast-wave formalism \citep{2006MNRAS.373.1074S} where cooling is delayed for $30$Myr to prevent spurious cooling. Additionally, stars inject thermal energy and metals into the surrounding ISM.
Star-formation is modelled according to the Kennicutt-Schmidt relation with a constant star-formation efficiency per free-fall time $\epsilon_\mathrm{ff}=0.1$ if the gas temperature is below $T=15000$K and the gas density is above $n=10.3 \mathrm{cm^{-3}}$ \citep{2013MNRAS.436..625S}.
The resulting galaxies range from dwarfs to Milky Way sized galaxies and reproduce a range of observational quantities.

\bsp	
\label{lastpage}
\end{document}